\pdfoutput=1

\documentclass[11pt,a4paper]{article}

\usepackage{jheppub}
\usepackage{amsmath}
\usepackage{multicol,bbm}
\usepackage{enumerate}
\usepackage{bibentry}

\usepackage{amssymb}
\usepackage{bm}
\usepackage{multirow}
\usepackage{caption}
\usepackage{tabularx}
\usepackage{enumitem}
\usepackage{extarrows}
\usepackage{tikz}
\usepackage{verbatim} % 多行注释

\graphicspath{{figures/}}  % 定义所有的 .eps 文件在 figures 子目录下

\def\be{\begin{equation}}
\def\ee{\end{equation}}
\def\ba{\begin{eqnarray}}
\def\ea{\end{eqnarray}}
\def\nl{\nonumber\\}

\def\a{\alpha}
\def\b{\beta}

\def\ad{{\dot\alpha}}
\def\bd{{\dot\beta}}

\def\l{\langle}
\def\r{\rangle}

\def\b#1{\overline{#1}}

\def\CP1{\mathbb{CP}^1}
\def\SL2C{\mathrm{SL}(2,\mathbb{C})}

\def\Z2{\mathbb{Z}_2}

\def\su2{{SU(2)}}

\def\a{{\alpha}}

\def\[{\left[}
\def\]{\right]}

\def\L{\Lambda}
\def\l{\lambda}
\def\lb{\tilde\lambda}

\def\a{\alpha}
\def\b{\beta}

\def\({\left(}
\def\){\right)}
\def\[{\left[}
\def\]{\right]}
\def\lan{\langle}
\def\ran{\rangle}
\def\<{\langle}
\def\>{\rangle}

\def\i2{\frac{i}{2}}

\def\2F1{\,_2{\rm F}_1}

\begin{document}

% Use the \preprint command to place your local institutional report
% number in the upper righthand corner of the title page in preprint mode.
% Multiple \preprint commands are allowed.
% Use the 'preprintnumbers' class option to override journal defaults
% to display numbers if necessary
%\preprint{}

%Title of paper
\title{Notes on Scattering Amplitudes as Differential Forms}

% repeat the \author .. \affiliation  etc. as needed
% \email, \thanks, \homepage, \altaffiliation all apply to the current
% author. Explanatory text should go in the []'s, actual e-mail
% address or url should go in the {}'s for \email and \homepage.
% Please use the appropriate macro foreach each type of information

% \affiliation command applies to all authors since the last
% \affiliation command. The \affiliation command should follow the
% other information
% \affiliation can be followed by \email, \homepage, \thanks as well.
%\author{}
%\email[]{Your e-mail address}
%\homepage[]{Your web page}
%\thanks{}
%\altaffiliation{}
%\affiliation{}
\author[a,b]{Song He}
\author[a,b]{,\,Chi Zhang}
\affiliation[a]{CAS Key Laboratory of Theoretical Physics, Institute of Theoretical Physics, Chinese Academy of Sciences, Beijing 100190, China}
\affiliation[b]{School of Physical Sciences, University of Chinese Academy of Sciences, No.19A Yuquan Road, Beijing 100049, China}
\emailAdd{songhe@itp.ac.cn}
\emailAdd{zhangchi@itp.ac.cn}
%Collaboration name if desired (requires use of superscriptaddress
%option in \documentclass). \noaffiliation is required (may also be
%used with the \author command).
%\collaboration can be followed by \email, \homepage, \thanks as well.
%\collaboration{}
%\noaffiliation

\date{\today}

\abstract{
Inspired by the idea of viewing amplitudes in ${\cal N}=4$ SYM as differential forms on momentum twistor space, we introduce differential forms on the space of spinor variables, which combine helicity amplitudes in any four-dimensional gauge theory as a single object. In this note we focus on such differential forms in ${\cal N}=4$ SYM, which can also be thought of as ``bosonizing" superamplitudes in non-chiral superspace. Remarkably all tree-level amplitudes in ${\cal N}=4$ SYM combine to a $d\log$ form in spinor variables, which is given by pushforward of canonical forms of Grassmannian cells. The tree forms can also be obtained using BCFW or inverse-soft construction, and we present all-multiplicity expression for MHV and NMHV forms to illustrate their simplicity. Similarly all-loop planar integrands can be naturally written as $d\log$ forms in the Grassmannian/on-shell-diagram picture, and we expect the same to hold beyond the planar limit. Just as the form in momentum twistor space reveals underlying positive geometry of the amplituhedron, the form in terms of spinor variables strongly suggests an ``amplituhedron in momentum space". We initiate the study of its geometry by connecting it to the moduli space of Witten's twistor-string theory, which provides a pushforward formula for tree forms in ${\cal N}=4$ SYM.
}

% insert suggested PACS numbers in braces on next line
%\pacs{}
% insert suggested keywords - APS authors don't need to do this
%\keywords{}

\maketitle
%must follow title, authors, abstract, \pacs, and \keywords
%\tableofcontents

\section{Introduction} \label{sec1}

The current work was inspired by a more intrinsic definition~\cite{Arkani-Hamed:2017vfh} of the amplituhedron for planar ${\cal N}=4$ super-Yang-Mills (SYM)~\cite{Arkani-Hamed:2013jha, Hodges:2009hk}: the key idea is that instead of thinking about scattering amplitudes merely as functions, they are to be thought of more fundamentally as differential forms on the space of physical kinematic data. In the context of planar ${\cal N} = 4$ SYM, the natural kinematic space is the space of momentum twistors $Z_i$ for the particles $i = 1, . . . , n$~\cite{Hodges:2009hk}. On this space the differential form has a natural purpose in life: it ``bosonizes" the super-amplitude (after stripping off the MHV tree amplitude) by treating the on-shell Grassmann variables $\eta_i$ as the differential of $Z_i$, $\eta_i \to d Z_i$. This seemingly innocuous move has dramatic geometric consequences: given a differential form, we can compute residues around singularities, which reveals the underlying positive geometry~\cite{Arkani-Hamed:2017tmz}. In particular for N$^{k{-}2}$MHV tree amplitude, we have a $d\log$ form of degree $4(k{-}2)$ in momentum twistor space, which encodes the geometry of the amplitudhedron. As discussed in~\cite{Arkani-Hamed:2017vfh}, the amplituhedron is given by the intersection of a top-dimensional region (with positive kinematics and correct winding related to $k$) and a family of $4(k{-}2)$-dimensional subspaces, and the form is completely specified by its behavior when “pulled back” to the subspace. On any such subspace, the form becomes the canonical form with logarithmic singularities on the boundaries of this positive geometry, from which locality and unitarity of the amplitudes emerge.

The same idea has played a crucial role in novel, geometric formulations for tree amplitudes in a wide range of massless theories~\cite{Arkani-Hamed:2017mur}, and for the cosmological polytope~\cite{Arkani-Hamed:2017fdk}.  For theories in general dimension, the natural kinematic space is the space of Mandelstam variables, and differential forms there, dubbed ``scattering forms", have very different meaning. For bi-adjoint $\phi^3$ theory, the scattering form is a $d\log$ form of degree $n{-}3$, which is given by the canonical form~\cite{Arkani-Hamed:2017mur} of an associahedron polytope in Mandelstam space, and the latter is the amplituhedron for this theory. Here the associahedron is again the intersection of a ``positive region" with $(n{-}3)$-dimensional subspaces, whose canonical form gives tree amplitudes of bi-adjoint scalar theory.  This picture of polytopes and associated canonical forms in Mandelstam space has been generalized to the so-called Cayley polytopes~\cite{Gao:2017dek,He:2018pue}, as well as the Halohedron which encodes one-loop amplitudes~\cite{Salvatori:2018aha}. The purpose of these scattering forms on Mandelstam space is that they encode color degrees of freedom, in general massless theories including Yang-Mills and non-linear sigma model~\cite{Arkani-Hamed:2017mur}. Wedge products of $d s$'s for tree-level cubic graphs satisfy the same Jacobi relations as color factors, and these scattering forms can be viewed as full color amplitudes without color factors~\cite{Arkani-Hamed:2017mur}. Of course for these theories, scattering forms cannot be interpreted as canonical forms of any positive geometry, since they are not $d\log$ forms anymore, but the idea of viewing color-dressed amplitudes as differential forms is useful {\it e.g.} in providing a ``geometric" origin of color/kinematic duality, double copy and connections to worldsheet formulation~\cite{Arkani-Hamed:2017mur}.

Given these advances, it is natural to ask what other kinematic space and amplitudes can we apply the general idea of differential forms, and what can we learn from it in the new context. In this paper we initiate the study of differential forms in the space of spinor-helicity variables, $\{\lambda_i, \tilde\lambda_i\}$, for massless amplitudes in four dimensions. The use of spinor-helicity variables is crucial for the drastic simplifications of massless amplitude starting from the celebrated work of Parke and Taylor~\cite{PhysRevLett.56.2459}. Instead of using polarization vectors, one computes amplitudes in helicity basis using spinor variables, which has nevertheless led to a proliferation of helicity amplitudes. It is thus highly desirable to combine all different helicity amplitudes into a single object. For supersymmetric theories, a nice way to do so is to package them as super-amplitudes using Grassmann variables in on-shell superspace({\it c.f.}~\cite{ArkaniHamed:2008gz}).  In sec.~\ref{sec2}, we propose a conceptually simple but powerful idea that also applies to general, non-supersymmetric theories, namely combining all helicity amplitudes in a {\it differential form} directly on kinematic space. For supersymmetric case, this amounts to ``bosonize" (non-chiral) super-amplitude with $\eta_i \to d\lambda_i, \tilde\eta_i \to d\tilde\lambda_i$, but we emphasize that such forms exist for supersymmetric as well as non-supersymmetric theories. As we will see shortly, this idea of combining helicity amplitudes into a single differential form is purely kinematical, which applies to any loop order in general massless theories with the highest spin not exceeding one. These include any gauge theories with massless fermions/scalars, such as (super-)Yang-Mills, massless QCD/QED {\it etc.}. To study the differential forms in full generality for any gauge theories goes beyond the scope of this paper; the idea can even be applied to the complete standard model, which deserves to be investigated on its own.

In a sense, these forms encoding helicity amplitudes of general theories are the analog of general scattering forms in Mandelstam space that encode color amplitudes. The analog of the $d\log$ form for bi-adjoint scalar theory here is the differential form which ``bosonize" super-amplitude in ${\cal N}=4$ SYM; it is remarkably a $d\log$ form in terms of spinor variables, which is our main focus. In sec.~\ref{sec3}, we will study in detail this form at tree level, which is the prototype for differential forms in the space of spinors. The existence of this $d\log$ form strongly suggests an ``amplituhedron" directly in momentum space, and it is still an important open question to determine completely this positive geometry. While encoding the same information, both the geometry and its canonical form are very different from those in momentum twistor space. In particular, the degree of the form is $2n{-}4$ for any tree amplitudes. The fact that these are $d\log$ forms immediately follows from the Grassmannian/on-shell diagram picture for ${\cal N}=4$ SYM. The differential form for each BCFW term of the superamplitude is the pushforward of canonical form of the corresponding cell of positive Grassmannian. In addition to the general proof, we study how to construct these forms though inverse-soft-construction, and work out explicitly all-$n$ MHV and NMHV forms as primary examples. As a byproduct, these remarkably simple formula for the forms provide results for non-chiral superamplitudes for ${\cal N}=4$ SYM~\cite{Huang:2011um,Plefka:2014fta}.

As we will discuss in sec.~\ref{sec4}, this clearly extends to any on-shell diagram or cell of Grassmannian, {\it i.e.} the form that ``bosonize" super-function of any on-shell diagram is the pushforward of the Grassmannian canonical form, including those needed for planar loop integrands. Conceptually, this provides a unified picture for tree and loops: an $L$-loop integrand in planar ${\cal N}=4$ SYM is a degree-$(2n{-}4{+}4L)$ differential form, which are interesting object themselves. Such forms can be obtained by using the all-loop recursion~\cite{ArkaniHamed:2010kv}, where the differential form of the additional pair of particles in the forward limit is turned into that of the additional loop variable. In practice, one can construct them using BCFW bridges familiar from on-shell diagrams.  Furthermore, these differential forms can be obtained even beyond the planar limit, and we provide evidences that they are $d\log$ forms in the full, non-planar ${\cal N}=4$ SYM~\cite{Arkani-Hamed:2014bca}.

The Grassmannian picture of $d\log$ form does not, however, provide an intrinsic definition of the amplitude form (as opposed to individual BCFW term {\it etc.}). The form must be completely determined by the underlying, still mysterious ``amplituhedron in momentum space"; to define the form as a single object, we give a ``twistor-string" pushforward formula for the tree amplitude form. Recall that Witten's twistor string theory~\cite{Witten:2003nn} for ${\cal N}=4$ SYM uses $G_+(2,n)$ as the moduli space, by summing over solutions of the Roiban-Spradlin-Volovich (RSV) scattering equations~\cite{PhysRevD.70.026009}, its top form pushes forward to exactly the full tree-level form. This simple observation provides strong hints for the underlying geometry, as we will show in sec.~\ref{sec5}. We end in sec.~\ref{sec6} with open questions.

\section{Scattering amplitudes as differential forms on %\textcolor{red}{momentum space}
\texorpdfstring{$\Gamma_n$}{TEXT}}  \label{sec2}

We will refer to the configuration space for $n$ massless momenta as $\Gamma_n$, where both $\lambda$'s and $\tilde\lambda$'s form  $2\times n$ matrices, and these two matrices are subject to momentum conservation:
\begin{align}\label{Gamma}
\Gamma_n=\{\L^\a, \tilde \L^\ad~|~\L^\a \cdot \tilde \L^\ad=0\}\equiv \{(\l^{\a}_1,\ldots, \l^\a_n), (\lb^{\ad}_1,\ldots, \lb^\ad_n)~|~\sum_{i=1}^n \l^{\a}_i\,\lb^{\ad}_i=0\}\,. % i=1,2,\ldots,n
\end{align}
The basic idea for combining helicity amplitudes as differential forms on $\Gamma_n$ is to dress each external state of helicity $h$ with a certain differential to cancel its little group weight. Recall that $| t \l, t^{-1} \lb \ran^{(h)} \to t^{-2h} | \l, \lb\ran^{(h)}$, thus this is possible for $|h|\leq 1$. For a massless particle with helicity $\pm 1$, it is natural to dress it with
\be
h=+1:\, (d \l)^2\equiv d\l^1 \wedge d\l^2=\frac 1 2 \epsilon_{\a \b} d\l^{\a} d\l^{\b}\ ;\quad
h=-1:\, (d\lb)^2\equiv d\lb^{\dot{1}} \wedge d\lb^{\dot{2}}=\frac 1 2 \epsilon_{\ad \bd} d\lb^{\ad} d\lb^{\bd}.\nonumber
\ee
This already allows one to combine different helicity amplitudes with gluons or photons, into a $2n$-form with no little-group weight. For example, the form for tree-level four-gluon amplitudes in Yang-Mills theory reads (here $s=(p_1+p_2)^2$ and $t=(p_2+p_3)^2$):
\be\label{4gluon}
{\cal F}_{\rm YM}^{\rm tree} (1,2,3,4)=\frac 1 {s\,t} \left(\lan 1\,2 \ran^2 [3\,4]^2 (d\lb_1)^2 (d \lb_2)^2 (d \l_3)^2 (d \l_4)^2 + {\rm perm.} \right)\,,
\ee
where ``perm." denotes terms with $(12)$ exchanged with $(13), (14), (23), (24), (34)$.
%for $n$-gluon amplitudes in Yang-Mills theory we have:
%\be
%{\cal F}_{n, {\rm YM}}=\cdots+\sum_{i,j} A^{\rm MHV}_{n, {\rm YM}} (\ldots, i^-,j^-, \ldots) (d\lb_i)^2 (d\lb_j)^2\, \prod_{a\neq i,j} (d\l_a)^2+\cdots\ ,
%\ee
%where we have suppressed contributions from helicity amplitudes except for those in the MHV sector, namely with two negative helicity gluons. Note that the definition is universal and applies to all loop orders, and an explicit example is the form for (color-ordered) four-gluon tree amplitudes:

For a massless fermion of helicity $-\frac{1}{2}$, it is natural to assign either $d \lb^{\ad}$ or $(d \lb)^2 d\l^{\a} $, while for helicity $+\frac 1 2$, either $d\l^{\a}$ or $(d\l)^2 d \lb^{\ad}$. One may worry that the unit does not work out uniformly, but for all theories we consider, the fermions always come in pair with $+\frac 1 2 $ and $-\frac 1 2$ helicities, so we can ``contract" a 1-form with a 3-form for a pair of fermions. It is thus natural to associate these spinor indices with the flavor indices of the fermions, and for a pair $(\bar{q}_i^-, q_j^+)$, we have four choices of 4-form to dress with:
\be
d\lb^{\ad}_i \wedge d\lb_{j, \ad}  (d\l_j)^2, \quad {\ad=1~{\rm or}~2}\ , \quad {\rm or} \quad (d \lb_i)^2 d\l^\a_i \wedge d\l_{j,\a}, \quad {\a=1~{\rm or}~2}\ .
%=(d \lb_i)^2 \frac 1 2 \epsilon_{\a,\b} (d \l_i)^{\a} (d \l_j)^{\b} \ ,\nonumber
\ee
with the indices lowered by $\epsilon_{\a \b}$ or $\epsilon_{\dot{\alpha}\dot{\beta}}$, and any of these choices can be used for a fermion flavor. %where $d\lb_i \wedge d\lb_j\equiv \frac 1 2 \epsilon_{\ad, \bd} d(\lb_i )^{\ad}  (d \lb_j)^{\bd}$ and similarly for $d\l_i \wedge d\l_j$.
Note that the 4-form for a pair of fermions have the same unit with the form for two spin-one particles. In this way, we can combine different amplitudes in massless QCD/QED, at least for theories with no more than four flavors of fermions, {\it e.g.} ${\cal F}^{\rm tree}_{n, {\rm QCD}} %=\sum_{k=2}^{n{-}2} {\cal F}^{\rm QCD~tree}_{n,k}
= {\cal F}^{\rm tree}_{n~{\rm gluons}}+ {\cal F}^{\rm tree}_{q \bar{q} +(n{-}2)~g}+ \cdots$, which contains the form for $n$-gluon tree amplitudes in Yang-Mills theory, that with $(n{-}2)$ gluons and a pair of quarks $q \bar{q}$, {\it etc.}.

Similar 4-forms can be used for a pair of scalars, since we will only consider theories where they are charged and always come in pair.  In general, any $n$-point amplitudes can be turned into a $2n$-form on $\Gamma_n$, which carries no little-group weight and combines different helicity amplitudes together as a single differential form. By definition, the form has only contractions among $\l$'s ($\lb$'s) and those among $d\l$'s ($d\lb$'s). This is certainly not the case for a general differential form on $\Gamma_n$, {\it i.e.} it can contain contractions like $\lan \l\, d\l \ran$ or $[ \lb\,d \lb]$, thus the $2n$ form we just defined is a special form on $\Gamma_n$, where we can safely distinguish between spinor indices of $\l$, $\lb$ and those of $d\l$, $d\lb$.

\subsection{Forms for supersymmetric amplitudes}

All the above results obviously apply to general theories, without any supersymmetry or special properties. However, if we do have a supersymmetry, it is very natural to associate the indices of $d\l^{\a}$ and $d\lb^{\ad}$ with the indices of R symmetry. In this way, the differential forms are in one-to-one correspondence with the superamplitudes, where we simply replace the anti-commuting variables $\eta$ or $\tilde\eta$ by the forms defined above. For example, with ${\cal N}=1$ supersymmetry, one can choose two superfield~\cite{Elvang:2011fx} $\Phi^{\dagger}=g^+ + \bar\psi_1 \tilde \eta^1 $, $\Phi= g_- + \psi'_1 \eta^1 $ with Grassmann variables $\eta^1$ and $\tilde\eta^1$. The rule is very clear: for $\Phi^{\dagger}$, we take $1\to (d\l)^2$ and $\tilde\eta^1 \to (d\l)^2 (d\lb)^1$, while for $\Phi$, $1\to (d\lb)^2$ and $\eta^1 \to (d\lb)^2 d\l^1$. Thus the superamplitude naturally leads to the form which packages all ${\cal N}=1$ helicity amplitudes.

In the following we will take ${\cal N}=4$ SYM as our primary example, where the spinor indices of the differential form $\a=1,2$, $\ad=1,2$ can be precisely associated with the $SU(2,2)$ R-symmetry indices. As mentioned above, $d \lb^{\ad}$ and $(d \lb)^2 d\l^{\a} $ can be associated with $2+2$ negative-helicity gluinos, which we denote as $\psi_{\ad}, \psi'_{\a}$, and similarly for positive-helicity ones, $\bar{\psi}_{\a}$ and $\bar{\psi}'_{\ad}$. The $6=1+4+1$ scalars, denoted as $\phi, \phi'_{\a, \ad}, {\phi}''$, are associated with $1, d\l ^{\a}  \wedge d\lb^{\ad}$, $(d\l)^2 (d\lb)^2$ respectively. The unit works out as before since we have pairs $\phi, \phi''$ or $\phi', \phi'$. The upshot is the expansion in analogous with the non-chiral superfield~\cite{Huang:2011um}
\begin{align}\label{exp}
\Phi(\l, \lb, d\l, d\lb)=&(d\l)^2 g^++ (d\lb)^2 g^-+\phi+ d\l^{\a} d \lb^{\ad} \phi'_{\a, \ad} + (d\l)^2 (d\lb)^2 \phi''\nl
&+ d\l^{\a} \psi_{\a} + (d\l)^2 d\lb^{\ad} \bar{\psi}_{\ad}+ d\lb^{\ad} \bar{\psi}'_{\ad}+ (d\lb)^2 d\l^{\a} \psi'_{\a} \,,
\end{align}
thus any $n$-point super-amplitude in ${\cal N}=4$ SYM, recast in a non-chiral $SU(2,2)$ form, can be translated into a $2n$-form with $\eta^{1,2} \to d\l^{1,2}$, $\tilde\eta^{1,2}\to (d\lb)^{1,2}$. %This is true for any N${}^{k{-}2}$MHV amplitudes, where all the terms are of the form $(d\lb)^{2k} (d\l)^{2(n{-}k)}$.

An important observation is that the $2n$-form obtained in this way vanishes identically for any parity-invariant, supersymmetric amplitudes. To see this, recall that for ${\cal N}$ supersymmetries, the superampliutde must contain overall fermionic delta functions $\delta^{\cal N}(Q) \delta^{\cal N}(\tilde Q)$, where $Q$ and $\tilde Q$ are $2{\cal N}$ supercharges which are parity conjugates of each other. Under the above prescription, the fermionic delta functions translate to a pair of ${\cal N}$ forms defined as follows:
\be
(d\,q)^{\a, \ad}:=\sum_{i=1}^n  \l_i^{\a}~(d\,\lb_i)^{\ad}\ , \quad (d\,\tilde q)^{\a, \ad}:=\sum_{i=1}^n (d \l_i)^{\a} \lb_i^{\ad}\ ,
\ee
where $\a=1,2$, $\ad=1,2$ allows one to have ${\cal N}=1,2,3,4$ supercharges $Q$'s and the same amount of $\tilde Q$'s. Thus, the $2n$ form must contain two overall factors $(d\,q)^{\cal N}$ and $(d\,\tilde q)^{\cal N}$, {\it e.g.} for ${\cal N}=4$ we have $(d\,q)^4:=\wedge_{\a=1}^2 \wedge_{\ad=1}^2\,d q^{\a ,\ad}$ and similarly $(d\,\tilde q)^4$ (with 3-pt amplitudes being the only exception). However, on the support of momentum conservation we have
\be
P=\sum_{i=1}^n \lambda_i \tilde\lambda_i=0~\implies~d q+ d \tilde q=d P=0~\implies~(d q)^{\cal N} \wedge (d\tilde q)^{\cal N}=0\ ,
\ee
thus the total form always vanishes for amplitudes ($n>3$) in any supersymmetric theories. This is not surprising since the vanishing of the form simply reflects the fact these amplitudes satisfy supersymmetry Ward identities. To extract the non-trivial part of the form, we need to strip off a copy of $(d q)^{\cal N}$, or equivalently up to a possible sign, $(d \tilde q)^{\cal N}$. For example, for the $2n$ form associated with amplitudes of ${\cal N}=4$ SYM, ${\cal F}_n^{{\cal N}=4}$, we define $2n{-}4$ form $\Omega_{n}$ and also $2n{-}8$ form $\hat{\Omega}_n$ (for $n>3$) by stripping off such overall forms:
\be
{\cal F}_n^{{\cal N}=4}:=(d q)^4\,\wedge\,\Omega_n^{{\cal N}=4}\ ,\qquad {\cal F}_{n>3}^{{\cal N}=4}:=(d q)^4\,\wedge\,(d \tilde q)^4\,\wedge\,\hat{\Omega}_n^{{\cal N}=4}\ ,
\ee
and similarly for $2n{-}{\cal N}$ form $\Omega^{(\cal N)}_n$ for ${\cal N}<4$ theories. Recall that for non-supersymmetric theories, the degrees of $d\l$'s and $d\lb$'s in ${\cal F}_n$ are unambiguous since it does not contain overall factors $d q=-d\tilde q$. It is determined by the MHV degree, $k$ (number of negative-helicity gluons for Yang-Mills amplitudes): ${\cal F}_n \sim (d\l)^{2(n{-}k)}~(d\lb)^{2k}$. For supersymmetric theories, ambiguities arise since we can swap $d q$ with $-d\tilde q$, and it is natural to go to $2n{-}2{\cal N}$ form which has no such overall factors. For example for $n>3$ we have $2\leq k\leq n{-}2$, and the $2n{-}8$ form for N${}^{k{-}2}$ MHV amplitudes in ${\cal N}=4$ reads: $\hat{\Omega}_n^{{\cal N}=4} \sim (d\l)^{2(n{-}k{-}2)}~(d\lb)^{2(k{-}2)}$.

\section{Differential forms for tree amplitudes of \texorpdfstring{$\mathcal{N}=4$}{TEXT} SYM} \label{sec3}

In this section, we study these tree-level differential forms, $\Omega^{{\cal N}=4}_{n,k}$, in detail. We start with $n=3$, with MHV and $\overline{\rm MHV}$ forms read:\begin{align}\label{3pt1}
&{\cal F}^{{\cal N}=4}_{3, {\rm MHV}}=(d\,q)^4 \Omega_{3,2}=\frac{(d\,q)^4\, (d \l_1 \lan 2\,3 \ran+ d \l_2 \lan 3\,1\ran+ d \l_3 \lan 1\,2\ran)^2}{\lan 1\,2\r~\lan 2\,3\r~\lan 3\,1\r}\,,\nl
&{\cal F}^{{\cal N}=4}_{3, \overline{\rm MHV}}=(d\,\tilde q)^4 \Omega_{3,1}=\frac{(d\,\tilde q)^4 \, (d \lb_1 [2\,3]+ d \lb_2 [3\,1]+ d \lb_3 [1\,2])^2}{[1\,2]~[2\,3]~[3\,1]}\,.
\end{align}
Note that although there is no $(dq)^{4} \wedge (d\tilde{q})^{4}$, $(dq)^{4}$ or $(d\tilde{q})^{4}$ vanishes by itself, since these $\lambda$'s or $\tilde{\lambda}$'s live in $G_{+}(2,3)$ which is two dimensional. The non-vanishing MHV/$\overline{\rm MHV}$ form is obtained by stripping off the overall 4-form, %and the first observation is that
the remaining 2-form is a $d\log$ form:
\begin{align}\label{3pt2}
\Omega_{3,2}=d\log \frac{\lan 1\,2\r}{\lan 3\,1\r} \wedge d\log \frac{\lan 2\,3\r}{\lan 3\,1\r}\,, \quad \Omega_{3,1}= d\log \frac{[1\,2]}{[3\,1]}\wedge d\log \frac{[2\,3]}{[3\,1]}\,.
\end{align}
This is not obvious {\it a priori}, but a direct use of the following identity %gives eq.(\ref{3pt2}):
\begin{equation*}
d\lambda_{i}\langle jk \rangle + d\lambda_{j}\langle ki \rangle + d\lambda_{k}\langle ij \rangle
= -(\lambda_{i}d\langle jk \rangle + \lambda_{j}d\langle ki \rangle +\lambda_{k}d\langle ij \rangle)\,.
\end{equation*}

Eq.(\ref{3pt2}) is the first example of our general claim: after stripping off an overall 4-form, the $2n{-}4$ form for any tree amplitude in ${\cal N}=4$ SYM is a $d\log$ form on $\Gamma_n$! This is not at all obvious from expressions of non-chiral superamplitudes, but as we show now it becomes manifest and is in fact guaranteed by the Grassmannian picture~\cite{ArkaniHamed:2012nw}. In this picture, any tree amplitude in ${\cal N}=4$ SYM is given by a sum of on-shell (super)-functions, which are in turn given by contour integrals over $C \in G_{k,n}$ ($C^{\perp} \in G(n{-}k, n)$ is its orthogonal complement):
\begin{equation}
f_{\gamma}^{(k)}=\oint_{C\in\gamma} \frac{d^{k\times n}C}{\operatorname{vol}\mathrm{GL}(k)}
\frac{\delta^{0\vert 2k}(C\cdot \tilde{\eta})\delta^{0\vert 2(n-k)}(C^{\bot}\cdot \eta)}{(1\cdots k)\cdots(n\cdots k-1)} \delta^{2k}(C\cdot \lambda) \delta^{2(n-k)}(C^{\bot} \cdot \tilde{\lambda})\,. \label{onshellfunc}
\end{equation}
Here each on-shell function is associated with a certain $(2n{-}4)$-dimensional cell $\gamma$ of the {\it positive} Grassmannian $G_{+}(k,n)$, which determines the contour; the measure here is the top form of $G_{k,n}$ but after the contour integral one is left with the canonical form of the $(2n{-}4)$-dim cell $\gamma$ ~\cite{ArkaniHamed:2012nw}. We have abbreviated the $2n{-}4$ bosonic and fermionic delta functions, and the latter are written in non-chiral, parity-invariant superspace (the dot product always means summing over particle index $a\cdot b:=\sum_{i=1}^{n}a_{i}b_{i}$).
We will see shortly that the differential form corresponding to any such on-shell function immediately turns into a $d\log$ form: it is in fact the pushforward of the canonical form on the cell $\gamma$ in $G_{+}(k,n)$.

To turn the super-function to a differential form, note that first there are four redundant bosonic delta functions $\delta^{4}(\lambda\cdot \tilde{\lambda})$ imposing momentum conservation and eight fermionic delta functions $\delta^{0|4} (Q) \delta^{0|4} (\tilde Q)$ (for $n>3$) for supercharge conservation. This can be seen by fixing {\it e.g.} the first two rows of $C$ to be $C_{\mu,a}=\l_a^{\alpha=\mu}$ or those of $C^{\perp}$ to be $C^{\perp}_{\tilde{\mu},a}=\lb_a^{\ad=\tilde\mu}$.
%with $Q^{\alpha, \dot{I}}=\sum_{a=1}^n \lambda_a^{\alpha} \tilde\eta_a^{\dot{I}}$ and similarly for $\tilde Q$.
As we have emphasized, to obtain the differential form, we need to strip off $\delta^4(P)$ and $\delta^{0|4} (\tilde Q)$ before making the replacement $\eta \to d\lambda$, $\tilde\eta \to d\tilde\lambda$. By fixing $C_{\mu=1,2,a}=\lambda_a^{\alpha=1,2}$ and pulling out $\delta^{4|4}(P|\tilde Q)$, the Jacobians for the bosonic and fermionic parts cancel each other. In this way, the differential form corresponding to the cell $\gamma$ reads (we denote the remaining part of $C$ as $C_{\mu', a}$ with $\mu'=3,\ldots, k$):
%$(d q)^4 \wedge (d \tilde q)^4=0$, thus either $\delta^{0|4} (Q)$ or $\delta^{0|4} (\tilde Q)$ needs to be removed before converting it to the form. It is obvious what we need to do: simply fix
\begin{align}
\Omega_{n,k}^{(\gamma)}=\int \omega_{n,k}^{(\gamma)}%\prod_{i=1}^{2n{-}4} \frac{d x_i}{x_i}
~\prod_{\mu'} \delta^2(C_{\mu'} (x) \cdot \tilde \lambda)  \prod_{\tilde{\mu}}
\delta^2(C^{\perp}_{\tilde{\mu}}(x) \cdot \lambda)
\bigwedge_{\mu'}(C_{\mu'} (x) \cdot d \tilde\lambda)^2\,\bigwedge_{\tilde{\mu}}(C_{\tilde{\mu}}^{\perp}(x)\cdot d\lambda)^2 \label{formGra1}
\end{align}
Let us define the $2n-4$ dimensional vector  $V_{i}:=(C_{\mu'}(x)\cdot\tilde\lambda^{\dot{\alpha}}, C^{\perp}_{\tilde{\mu}} (x)\cdot \lambda^{\alpha})$ where $i=\{(\mu', \dot\alpha), (\tilde{\mu}, \alpha)\}$ with $\mu'=3,\ldots,k, \tilde\mu=1,\ldots, n{-}k$ and $\a, \ad=1,2$ (thus $i=1,2,\ldots,2n-4$). Note that the delta functions imply $V_i=0$ and hence $d\,V_i=0$, thus
\begin{align}\label{identity1}
0=\sum_j \frac{\partial V_i}{\partial x_j} d x_{j} + \delta_{i,(\mu',\dot\alpha)}\,
C_{\mu'} \cdot d \tilde\lambda^{\dot{\alpha}}+\delta_{i, (\tilde{\mu}, \alpha)} \,
C_{\tilde{\mu}}^{\perp}\cdot  d \lambda^{\alpha}  \:,
\end{align}
which can be decomposed into two parts
\begin{equation}
\sum_j \frac{\partial V_{i=(\mu', \ad)} }{\partial x_j} d x_j =- C_{\mu'}\cdot  d\tilde\lambda^{\dot{\alpha}}\quad {\rm and} \quad \sum_j \frac{\partial V_{i=(\tilde\mu, \a)} }{\partial x_j} d x_j =- C_{\tilde{\mu}}^{\perp} \cdot d \lambda^{\alpha} \,. \label{identity2}
\end{equation}
By taking the wedge product of $2n{-}4$ copies of eq.\eqref{identity2}, we obtain
\begin{equation}
d x_1 \wedge \cdots \wedge d x_{2n{-}4}~\det (\{ \partial V_i/\partial x_j\})
=\bigwedge_{\mu'} (C_{\mu'}(x) \cdot d \tilde\lambda)^2\,\bigwedge_{\tilde{\mu}}(C_{\tilde{\mu}}^{\perp}(x)\cdot d\lambda)^2\,. \label{identity3}
\end{equation}
Note that the Jacobian is exactly that comes from integrating over the delta functions in \eqref{formGra1}, and the remaining form is nothing but the right hand side of eq.(\ref{identity3}). Thus, after solving the $x$ variables via the delta functions, we find the final result is
\begin{align}\label{formGra2}
\Omega_{n,k}^{(\gamma)}= \frac{d x_1(\Lambda, \tilde\Lambda) \wedge \cdots \wedge d x_{2n{-}4} (\Lambda, \tilde\Lambda)}{  x_1 (\Lambda, \tilde\Lambda) \cdots x_{2n{-}4} (\Lambda, \tilde\Lambda)}=\omega_{n,k}^{(\gamma)}(\Lambda, \tilde\Lambda)\ ,
\end{align}
where the differential hits $\l$'s and $\lb$'s. Thus, it defines a pushforward from the canonical form $\omega_{n,k}^{(\gamma)}$ on the cell $\gamma$ of Grassmannian space to the differential form $\Omega_{n,k}^{(\gamma)}$ on the $\Gamma_{n}$

Let us see how our formalism works for the cases of MHV/$\overline{\mathrm{MHV}}$ and 6pt NMHV.

\paragraph{MHV/$\overline{\mathbf{MHV}}$ cases.} In the case of MHV, the canonical form on the Grassmannian's side is simply the canonical form for the $G_{+}(2,n)$ (top cell)
\begin{equation*}
\omega_{n,2} = \frac{d^{2n}C}{\operatorname{vol}\mathrm{GL}(2)(12)(23)\cdots (n1)} \:,
\end{equation*}
and we can take the solution to be  $C_{\alpha a}=\lambda_{a}^{\alpha}$. Now, to put the differential form into the form of $d\log$, we can make a gauge fixing by acting with a particular $GL(2)$:
\begin{align*}
C^{\ast} &= \frac{1}{\langle 12\rangle }\left(
\begin{array}{cc}
\lambda _{2}^{2} & -\lambda _{2}^{1} \\
-\lambda _{1}^{2} & \lambda _{1}^{1}%
\end{array}%
\right) \left(
\begin{array}{cccc}
\lambda _{1}^{1} & \lambda _{2}^{1} & \cdots & \lambda _{n}^{1} \\
\lambda _{1}^{2} & \lambda _{2}^{2} & \cdots & \lambda _{n}^{2}%
\end{array}%
\right) \\
&=\frac{1}{\langle 12\rangle }\left(
\begin{array}{ccccccc}
\langle 12\rangle & 0 & \langle 32\rangle & \cdots & \langle i2\rangle &
\cdots & \langle n2\rangle \\
0 & \langle 12\rangle & \langle 13\rangle & \cdots & \langle 1i\rangle &
\cdots & \langle 1n\rangle%
\end{array}%
\right)
\end{align*}%
such that by evaluating $\omega_{n,k=2}\rvert_{C^{\ast}}$ we immediately obtain a $d\log$ form
\begin{equation}
\Omega^{\rm tree}_{n,2} =\prod_{i=1}^{n}\left( \frac{%
\langle i\, i+1\rangle }{\langle 12\rangle }\right) ^{-1}\prod_{i=3}^{n}d\frac{%
\langle i2\rangle }{\langle 12\rangle }\wedge d\frac{\langle 1i\rangle }{%
\langle 12\rangle } = \prod_{i=2}^{n-1} d\log \frac{\langle i\,i+1
\rangle}{\langle 1\,i+1\rangle} \wedge  d\log \frac{\langle 1\,i+1
\rangle}{\langle 1\,2\rangle} \:. \label{MHVspecial}
\end{equation}
The result for $\overline{\mathrm{MHV}}$ is the same with all $\lambda$'s replaced with $\tilde{\lambda}$'s. In particular, let's record the differential form for $n=4$ (MHV and $\overline{\mathrm{MHV}}$ coincide here):
\begin{align}\label{4pt}
\Omega_{4,2}^{\rm{tree}} &= \frac{ (d\,\tilde q)^4}{s t}=d\log \frac{\lan 1\,2\ran}{\lan 1\,3\ran}\wedge d\log \frac{\lan 2\,3\ran}{\lan 1\,3\ran}\wedge d\log \frac{\lan 3\,4\ran}{\lan 1\,3\ran} \wedge d\log \frac{\lan 4\,1\ran}{\lan 1\,3\ran}\nl
&= \frac{ (d\,q)^4}{s t}=d\log \frac{[1\,2]}{[2\,4]} \wedge d\log \frac{[2\,3]}{[2\,4]} \wedge d\log \frac{[3\,4]}{[2\,4]} \wedge d\log \frac{[4\,1]}{[2\,4]} \:,
%\wedge_{i=1}^4\, d \log x_i=\wedge_{i=1}^4 d\log \bar{x}_i\,,
\end{align}
where the equality between MHV and $\overline{\rm{MHV}}$ forms follows from momentum conservation.

In fact, it is well known that (\ref{MHVspecial}) is one of many equivalent representations of the canonical form of  $G_+(2,n)$. To write a general one, we use the terminology of the cluster algebra for $G_+(2,n)$. Recall that here each cluster correspond to a triangulation of $n$-gon, and for MHV case we label the $n$ edges as $\langle i\, i{+}1\rangle$ for $i=1,2,\cdots, n$, and the $n{-}3$ diagonals of the triangulation as $\langle i\,j\rangle$ (connecting vertex $i$ and $j$). Given such $2n{-}3$ brackets, one can then take {\it any} $2n{-}4$ independent ratios of them, which we call $x_1, x_2, \cdots, x_{2n{-}4}$. For example, one can choose the triangulation such that the $2n{-}3$ edges correspond to $\lan i\,i{+}1 \ran$ for $i=1,\ldots, n$ and $\lan 1 i\ran$ for $i=3,\ldots, n{-}1$, and one choice of $x$'s can be given by the $2n{-}4$ ratios in (\ref{MHVspecial}). A simple but remarkable fact is that the wedge product of these $2n{-}4$ $d\log x$'s are independent of the triangulation (up to a sign). Thus the differential form for MHV amplitude can be written for any cluster of $\Lambda\in G_+(2,n)$:
\begin{equation}
\Omega^{\rm tree}_{n, {\rm MHV}}=d\log x_1 \wedge d\log x_2 \wedge \cdots \wedge d \log x_{2n{-}4}\ ,
\label{MHV}
\end{equation}
Similarly for $\overline{\mathrm{MHV}}$ we define $\bar{x}$'s as ratios of square brackets, for any cluster $\tilde\Lambda\in G_+(2,n)$:
\be
\Omega^{\rm tree}_{n, \overline{\rm MHV}}=d\log {\bar x}_1\wedge d\log {\bar x}_2 \wedge \cdots \wedge d\log {\bar x}_{2n{-}4}\ .
\ee
To prove MHV or $\overline{\mathrm{MHV}}$ tree forms are independent of the triangulation of a $n$-gon (up to a sign), it is sufficient to prove this holds for $n=4$. This is because that if the differential form for $4$-pt MHV tree is independent of the triangulation, then we can do the so-called \emph{flip} operation, which transform a triangulation of a tetragon into the other, for any tetragon in this triangulated $n$-gon. It is obvious that any two triangulations of a $n$-gon can be transformed to each other by a sequence of flip operations. On the other hand, we have
\begin{equation*}
d\log\frac{\langle 2\,3\rangle}{\langle 1\,2\rangle}\wedge d\log\frac{\langle 3\,4\rangle}{\langle 1\,2\rangle}\wedge d\log\frac{\langle 4\,1\rangle}{\langle 1\,2\rangle}\wedge d\log\frac{\langle 1\,3\rangle}{\langle 1\,2\rangle}
=-d\log\frac{\langle 2\,3\rangle}{\langle 1\,2\rangle}\wedge d\log\frac{\langle 3\,4\rangle}{\langle 1\,2\rangle}\wedge d\log\frac{\langle 4\,1\rangle}{\langle 1\,2\rangle}\wedge d\log\frac{\langle 2\,4\rangle}{\langle 1\,2\rangle}\,.
\end{equation*}
%since
%\begin{equation*}
%d\log\frac{\langle 1\,3\rangle}{\langle 1\,2\rangle}+d\log\frac{\langle 2\,4\rangle}{\langle 1\,2\rangle}
%=d\log\left(\frac{\langle 3\,4\rangle}{\langle 1\,2\rangle}+\frac{\langle 1\,4\rangle}{\langle 1\,2\rangle}\frac{\langle 2\,3\rangle}{\langle 1\,2\rangle}\right) \:.
%\end{equation*}

\paragraph{6pt NMHV.} Beyond the MHV/$\overline{\mathrm{MHV}}$ cases, the tree form becomes more interesting since the corresponding Grassmannian cell is no longer the top cell of $G_{+}(k,n)$. Instead, we have a sum of $(2n{-}4)$-dim cells which are dictated by {\it e.g.} BCFW representation of tree amplitudes. The first such example is 6-pt NMHV amplitude, where each BCFW term comes from a $8$-dim cell of $G_{+}(3,6)$. There are six such cells $\gamma_{i}$ with $i=1,\ldots,6$, each of which corresponds to the $C$ matrix with vanishing minor $(i, i{+}1, i{+}2)$~\cite{ArkaniHamed:2012nw}. Note that $\gamma_i$'s are related to each other by cyclic shifts, and since the corresponding cell in $C^{\perp}$ have $(5{-}i, 6{-}i, 7{-}i)=0$ the on-shell functions for $\gamma_i$ and $\gamma_{5{-}i}$ are related by parity. It is well known that there are two BCFW representations, related to each other by parity, for 6-pt NMHV tree amplitudes. Therefore, the tree form can be written as a sum of three terms in either of the representations:
\be\label{6ptNMHV}
\Omega_{6,3}^{\rm tree}=\Omega_{6,3}^{\gamma_1}+ \Omega_{6,3}^{\gamma_3}+ \Omega_{6,3}^{\gamma_5}=\Omega_{6,3}^{\gamma_4}+ \Omega_{6,3}^{\gamma_2}+ \Omega_{6,3}^{\gamma_6}\ .
\ee
It is straightforward to work out the form for cell $\gamma_1$ (others are obtained by cyclic shift):
\be\label{6ptNMHV2}
\Omega_{6,3}^{\gamma_1}=\frac{(d\tilde{q})^4(d\l_1\lan 2\,3\ran + d\l_2\lan 3\,1\ran+d\l_3\lan 1\,2\ran)^2 (d\lb_4 [5\,6] + d\lb_5 [6\,4] +d\lb_6 [4\,5] )^2}{s_{123}\,\lan 1\,2\ran\,\lan 2\,3\ran\,[4\,5]\,[5\,6]\,\lan 1|5{+}6|4]\,\lan 3|4{+}5|6]} \:,
\ee
%\nl&=d\log (x_1)\wedge d\log (x_2) \wedge \ldots \wedge d\log (x_8)\ ,\ea
since in this case we have the solution
\begin{equation*}
C^{\ast}= \begin{pmatrix}
\lambda_{1}^{1} & \lambda_{2}^{1} &\lambda_{3}^{1} &\lambda_{4}^{1} & \lambda_{5}^{1} & \lambda_{6}^{1} \\
\lambda_{1}^{2} & \lambda_{2}^{2} &\lambda_{3}^{2} &\lambda_{4}^{2} & \lambda_{5}^{2} & \lambda_{6}^{2} \\
0 & 0 & 0 & [56] & [64] & [45]
\end{pmatrix}
\end{equation*}
with vanishing minor $(1,2,3)$, %and satisfying the equation $C\cdot\tilde{\lambda} =0$,
where the numerator arises from $(C^{\ast}\cdot d\tilde\lambda)^{2} (C^{\ast\perp}\cdot d\lambda)^{2}$ after stripping off the factor $(dq)^{4}$ and the denominator is the product of five non-vanishing minor $(i,i+1,i+2)$ of $C^{\ast}$ with $i=2,\ldots,6$.
%\nl&=d\log (x_1)\wedge d\log (x_2) \wedge \ldots \wedge d\log (x_8)\ ,\ea
It is equivalent to $d\log (x_1)\wedge d\log (x_2) \wedge \cdots \wedge d\log (x_8)$ with one choice of canonical variables $x_1,x_2, \ldots, x_8$
\begin{equation*}
x_1=\frac{\lan 1\,2\ran}{\lan 3\,1\ran}\ ,\ x_2=\frac{\lan 2\,3\ran}{\lan 3\,1\ran}\ ,\
x_3=\frac{[\widehat{3}\,4]}{[ \widehat{3}\,\widehat{1}]}\ ,\  x_4=\frac{[4\,6]}{[\widehat{3}\,\widehat{1}]}\ ,\
x_5=\frac{[6\,\widehat{1}]}{[\widehat{3}\,\widehat{1}]}\ ,\ x_6=\frac{[\widehat{1}\,4]}{[\widehat{3}\,\widehat{1}]}\ ,
x_7=\frac{[5\,4]}{[6\,4]}\ , x_8=\frac{[6\,5]}{[6\,4]}\ .
\end{equation*}
which also can be written as a wedge product of a 3-pt MHV form and a 5-pt $\overline{\mathrm{MHV}}$ from
\begin{equation}
\Omega_{6,3}^{\gamma_1}= \Omega_{k=3}(\widehat{1},\widehat{3},4,5,6)\wedge \Omega_{k=2}(1,2,3) \label{6ptISF}
\end{equation}
where hats mean that these momenta are shifted by
\begin{equation*}
\tilde \lambda_{\hat{1}}=\tilde{\lambda}_{1}+\frac{\lan 1\,2\ran}{\lan 1\,3\ran} \tilde\lambda_{2}\:, \qquad
\tilde \lambda_{\hat{3}}=\tilde{\lambda}_{3}+\frac{\lan 2\,3\ran}{\lan 1\,3\ran} \tilde\lambda_{2} \:.
\end{equation*}
The structure exhibited in eq.\eqref{6ptISF} implies another construction of differential forms besides pushforward map, which is called Inverse-soft (IS) construction and described in detail below. This method allow us to build the differential form for any tree amplitude recursively instead of working out the pushforward map.

\subsection{Inverse-soft construction for the tree form}

As we have seen in the above, one way to obtain the differential forms is the pushforward from a certain cell of positive Grassmannian. Such cells also have a diagrammatic representation, that are on-shell diagrams~\cite{ArkaniHamed:2012nw}.
A particularly interesting procedure to build on-shell diagrams, is by attaching the so-called inverse soft factor, which here amounts to attaching a 2-form to the $(2n{-}4)$-form for $n$ points to obtain the $(2n{-}2)$-form for $n{+}1$ points.  Attaching $k$-preserving and $k$-increasing inverse soft factor correspond to inserting two cases of \eqref{3pt2} respectively:
\begin{equation}
\begin{split}
k-{\rm preserving}:\quad &\Omega(\ldots, n, n{+}1, 1)=\Omega(\ldots, \hat{n}, \hat{1}) \wedge d\log \frac{\lan n\,n{+}1 \ran}{\lan n\,1\ran} \wedge d\log \frac{\lan n{+}1\,1 \ran}{\lan n\,1\ran}\,, \\
k-{\rm increasing}:\quad &\Omega(\ldots, n, n{+}1, 1)=\Omega(\ldots, \hat{n}, \hat{1}) \wedge d\log \frac{[n\,n{+}1 ]}{[ n\,1 ]} \wedge d\log \frac{[ n{+}1\,1 ]}{[n\,1]}\, ,
\end{split}   \label{ISF}
\end{equation}
where the first case $\tilde \lambda_{\hat{n}}=\tilde\lambda_n+\frac{\lan n{+}1\,1\ran}{\lan n\,1\ran} \tilde\lambda_{n{+}1}$, $\tilde\lambda_{\hat{1}}=\tilde\lambda_1+\frac{\lan n\,n{+}1\ran}{\lan n\,1\ran} \tilde\lambda_{n{+}1}$ ($\lambda$'s unchanged) and the second case $\lambda_{\hat{n}}=\lambda_n+\frac{[ n{+}1\,1 ]}{[n\,1]}\lambda_{n{+}1}$, $\lambda_{\hat{1}}=\lambda_1+\frac{[n\,n{+}1 ]}{[ n\,1 ]} \lambda_{n{+}1}$ ($\tilde\lambda$'s unchanged). In the context of on-shell diagrams, these two operations correspond to
\begin{equation*}
\raisebox{-51pt}{\includegraphics[scale=0.7]{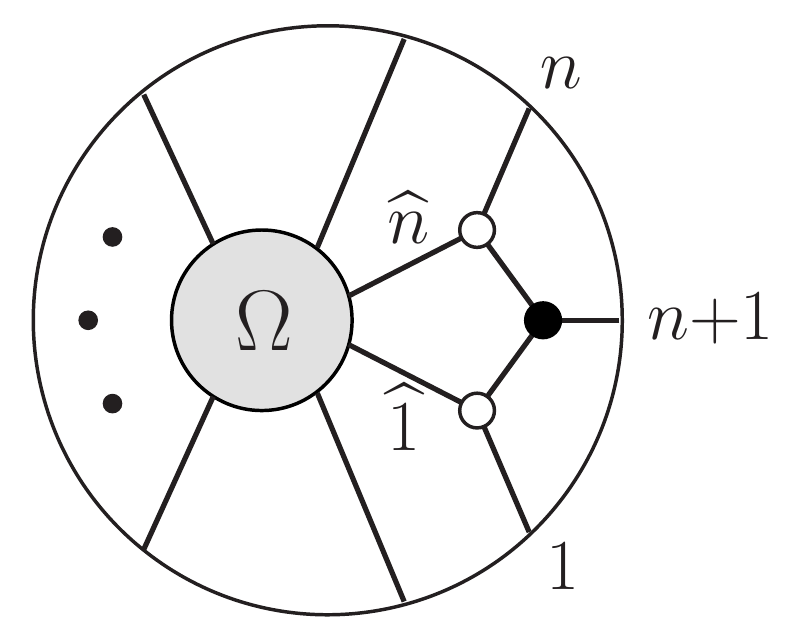}} \qquad \qquad
\raisebox{-51pt}{\includegraphics[scale=0.7]{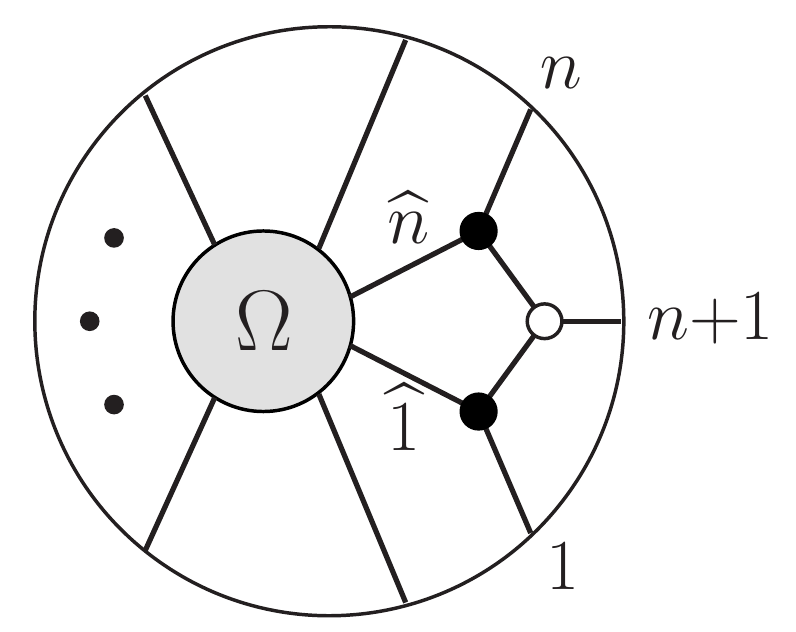}}
\end{equation*}
These operations actually are two special cases of BCFW constructions: attaching $k$-preserving inverse-soft factor correspond to the BCFW construction $(\bullet\otimes \mathcal{A}_{3}^{(1)})$ and  attaching $k$-preserving inverse-soft factor correspond to $(\mathcal{A}_{3}^{(2)}\otimes \bullet)$. Since the $n$-pt MHV amplitude $\mathcal{A}_{n}^{(2)}$ has the BCFW-recursion $(\mathcal{A}_{n-1}^{(2)}\otimes \mathcal{A}_{3}^{(1)})$, we arrive at the n-pt MHV form (\ref{MHV}) again by repeated use of this recursion, and similarly for the $\overline{\mathrm{MHV}}$ cases. It is not difficult to include the contribution (\ref{6ptNMHV2}) to the 6-pt NMHV form in the IS construction by recognizing the particle 5 as the $k$-increasing inverse-soft-factor or particle 2 as the $k$-preserving inverse-soft-factor, since (\ref{6ptNMHV2}) has two equivalent forms
\begin{align}
\Omega_{6,3}^{\gamma_1}&= \Omega_{k=2}(1,2,3,\widehat{4},\widehat{6})\wedge d\log \frac{[4\,5]}{[6\,4]}
\wedge d\log \frac{[5\,6]}{[6\,4]} \\
&=\Omega_{k=3}(\widehat{1},\widehat{3},4,5,6)\wedge d\log\frac{\langle 1\,2\rangle}{\langle3\,1\rangle}\wedge
d\log\frac{\langle 2\,3\rangle}{\langle3\,1\rangle} \:,
\end{align}
where the second equality can be shown by using of eq.(\ref{4pt}).

If an on-shell diagram can be obtained by successively adding particles to a 3-pt on-shell diagram, such an on-shell diagram is called `\emph{inverse-soft constructible}'. It turns out that all on-shell diagrams up to 13 particles are inverse-soft constructible. There are some contributions to the 14-pt $\mathrm{N}^{5}\mathrm{MHV}$ which are not inverse-soft constructible, for instance, the on-shell diagram whose BCFW-decomposition is $((\mathcal{A}_{3}^{(2)}\otimes(\mathcal{A}_{4}^{(2)}\otimes\mathcal{A}_{4}^{(2)}))\otimes\mathcal{A}_{3}^{(1)})
\otimes(\mathcal{A}_{3}^{(2)}\otimes((\mathcal{A}_{4}^{(2)}\otimes\mathcal{A}_{4}^{(2)})\otimes\mathcal{A}_{3}^{(1)}))
$.\cite{ArkaniHamed:2012nw} Fortunately, the on-shell diagrams that have the contribution to an amplitude are depend on the BCFW recursion scheme\cite{ArkaniHamed:2012nw}, and there is indeed a scheme $\{-2,2,0\}$ so that all on-shell diagrams for any tree amplitude are inverse-soft constructible.

Based on the above discussion, an effective method to construct the differential forms for tree amplitudes is the IS construction. If the IS decomposition of a on-shell diagram is found, then as is its differential form. There is a effective way to obtain the IS decomposition via the permutation labelling a on-shell diagram, if a permutation $\sigma$ for some particle $a$ has $\sigma(a-1)=a+1$ or $\sigma(a+1)=a-1$, then $a$ is a k-preserving or k-increasing inverse soft-factor, respectively. For instance, the permutation corresponding to (\ref{6ptNMHV2}) is $\{3,5,6,7,8,10\}$, then it is not difficult to recognize $2$ as a k-preserving inverse soft-factor and $5$ as a k-increasing inverse soft-factor. After removing this inverse soft-factor, the permutation becomes:
\begin{itemize}[leftmargin=2em]
\item[(i)] if $a$ is a k-preserving inverse soft-factor, then taking $\sigma(a-1)$ equals to $\sigma(a)$, finding the particle whose permutation equals to $a$ and changing its permutation to $a+1$, leaving the other particles unchanged.
\item[(ii)] if $a$ is a k-increasing inverse soft-factor, then taking $\sigma(a+1)$ equals to $\sigma(a)$, finding the particle whose permutation equals to $a$ and changing its permutation to $a-1$, leaving the other particles unchanged.
\end{itemize}
Here we illustrate how this procedure work by the following 8-pt $\mathrm{N}^{2}\mathrm{MHV}$ BCFW term
\begin{equation}
\{3,6,7,8,10,12,9,13\} \xlongrightarrow{\text{remove 2}} \{6,7,8,11,12,9,13\} \xlongrightarrow{\text{remove 8}}
\{6,7,9,11,12,13\} \:,
\end{equation}
where \{6,7,8,11,12,9,13\} is the permutation of \{1,3,4,5,6,7,8\} and \{6,7,9,11,12,13\} is the permutation of \{1,3,4,5,6,7\}. The last permutation correspond to the 6-pt $\overline{\mathrm{MHV}}$ amplitude, that is why we stop there. In the first step, we recognize $2$ as a k-preserving inverse soft-factor, then we change $\sigma(1)$ from 3 to $\sigma(2)=6$, since $\sigma(5)=10\:(2\operatorname{mod} 8)$, then $\sigma(5)$ equals to 11 in the new permutation, similarly for the second step. (Note that the second permutation is the permutation of \{1,3,4,5,6,7\}, we need to relabel these particles by \{1,2,3,4,5,6,7\} for repeating this procedure, after this procedure has been done, we can recover the labels of these particles.) We can represent this decomposition by such a formula
\begin{equation}
\{3,6,7,8,10,12,9,13\} = B(1,3,4,5,6,7)\otimes W(7,8,1) \otimes W(1,2,3) \:,
\end{equation}
where $B$ denotes the $\overline{\mathrm{MHV}}$ amplitude and $W$ denotes the MHV amplitude, this decomposition is order-dependent except some special cases. In what follows, we will give some further details of this decomposition and introduce a diagrammatic representation.

In the case of MHV or $\overline{\mathrm{MHV}}$, we find the $x$-variables correspond to edges and non-crossing diagonals of a $n$-gon, thus we represent the $n$-pt MHV or $\overline{\mathrm{MHV}}$ differential form by a white $n$-gon or a black $n$-gon, respectively. With the triangulation of this $n$-gon specified, the differential form is determined at the same time, for example,  a $n$-pt MHV differential form with $x$-variables fixed can be represented by
\begin{equation}
\raisebox{-1.3cm}{
\begin{tikzpicture}[scale=0.5]
\draw[line width=.03cm]  (0,0) coordinate (1) node [left=0pt]{1}
  --++(135:2) coordinate (2) node [left=0pt]{2}
  --++(90:2) coordinate (3) node [left=0pt]{3}
  --++(45:2) coordinate (4) node [left=0pt]{4}
  --++(0:2) coordinate (5) node [right=0pt]{5}
  --++(-45:2) coordinate (6) node [right=0pt]{6}
  ++(-90:2) coordinate (7) node [right=0pt]{$n-1$}
  --++(-135:2) coordinate (8) node [right=0pt]{$n$}
  --+(180:2);
  \draw[line width=.03cm,dashed] (6)--(7);
  \draw[line width=.02cm] (1)--(3);
  \draw[line width=.02cm] (1)--(4);
  \draw[line width=.02cm] (1)--(5);
  \draw[line width=.02cm] (1)--(6);
  \draw[line width=.02cm] (1)--(7);
\end{tikzpicture}}
= \prod_{i=2}^{n-1} d\log \frac{\langle1\,i\rangle}{\langle i+1\, 1\rangle}\wedge d\log \frac{\langle i\, i+1\rangle}{\langle i+1\, 1\rangle} \:.
\end{equation}
This representation can be easily generalized to non-MHV($\overline{\mathrm{MHV}}$) cases via inverse-soft construction. In the spirit of the above construction, each inverse-soft factor with its two adjacent particles (for each step of IS construction) can be viewed as the building block of this diagrammatic representation and represented by a white triangle or black triangle respectively for the k-preserving case and k-increasing case. For example, those three contributions to 6-pt NMHV amplitude are represented by
\begin{equation*}
\raisebox{0cm}{
\begin{tikzpicture}[scale=.5]
   \draw[line width=.03cm](0,0) coordinate (1) node[left=0pt]{$\widehat{1}$}
  --++(120:2) coordinate (2) node[left=0pt]{2}
  --++(60:2) coordinate (3) node[left=0pt]{$\widehat{3}$}
  --++(0:2) coordinate (4) node[right=0pt]{4}
  --++(-60:2) coordinate (5) node[right=0pt]{5}
  --++(-120:2) coordinate (6) node[right=0pt]{6}
  --+(180:2) ;
     \fill[black!20] (1)--(3)--(4)--(5)--(6)--(1);
     \draw[line width=.03cm] (1)--(3);
\end{tikzpicture}
} \qquad
\raisebox{0cm}{
\begin{tikzpicture}[scale=.5]
   \draw[line width=.03cm](0,0) coordinate (1) node[left=0pt]{1}
  --++(120:2) coordinate (2) node[left=0pt]{2}
  --++(60:2) coordinate (3) node[left=0pt]{$\widehat{3}$}
  --++(0:2) coordinate (4) node[right=0pt]{4}
  --++(-60:2) coordinate (5) node[right=0pt]{$\widehat{5}$}
  --++(-120:2) coordinate (6) node[right=0pt]{6}
  --+(180:2) ;
     \fill[black!20] (1)--(2)--(3)--(5)--(6)--(1);
     \draw[line width=.03cm] (3)--(5);
\end{tikzpicture}
} \qquad
\raisebox{0cm}{
\begin{tikzpicture}[scale=.5]
   \draw[line width=.03cm](0,0) coordinate (1) node[left=0pt]{$\widehat{1}$}
  --++(120:2) coordinate (2) node[left=0pt]{2}
  --++(60:2) coordinate (3) node[left=0pt]{3}
  --++(0:2) coordinate (4) node[right=0pt]{4}
  --++(-60:2) coordinate (5) node[right=0pt]{$\widehat{5}$}
  --++(-120:2) coordinate (6) node[right=0pt]{6}
  --+(180:2) ;
     \fill[black!20] (1)--(2)--(3)--(4)--(5)--(1);
     \draw[line width=.03cm] (5)--(1);
\end{tikzpicture}
}
\end{equation*}
which correspond to $\gamma_{1}$, $\gamma_{3}$ and $\gamma_{5}$, respectively. Actually, for n-pt NMHV case, the differential form is represented by
\begin{align}
\sum_{1<i<j<n}
\raisebox{-1.3cm}{
\begin{tikzpicture}[scale=0.2]
   \draw[line width=.03cm]  (0,0) coordinate (12) node [left=0.2pt,below=0.1pt]{\scriptsize$n$}
  --++(150:2) coordinate (1) node [left=0pt]{\scriptsize$\widehat{1}$}
  --++(120:2) coordinate (2)
  ++(90:2) coordinate (3)
  --++(60:2) coordinate (4) node [above=0.3pt]{\scriptsize$\widehat{i-1}\:\:$}
  --++(30:2) coordinate (5) node [left=0.1pt,above=0.1pt]{\scriptsize$\widehat{i}$}
  --++(00:2) coordinate (6)
  ++(-30:2) coordinate (7)
  --++(-60:2) coordinate (8) node [right=0pt,]{\scriptsize$\widehat{j}$}
  --++(-90:2) coordinate (9) node [right=0pt]{\scriptsize$\widehat{j+1}$}
  --++(-120:2) coordinate (10)
  ++(-150:2) coordinate (11)
  --+(180:2);
  \fill[black!20] (1)--(4)--(5)--(8)--(9)--(1);
  \draw[line width=.02cm] (1)--(4)--(5)--(8)--(9)--(1);
  \draw[line width=.03cm,dashed] (2)--(3);
  \draw[line width=.03cm,dashed] (6)--(7);
  \draw[line width=.03cm,dashed] (10)--(11);
\end{tikzpicture}
}&=\sum_{1<i<j<n} \Omega_{k=2}(1,\cdots ,i-1)\wedge  \Omega_{k=2}(i,\cdots,j)\wedge \Omega_{k=2}(j+1,\cdots ,n,1)
\nonumber \\
&\quad \wedge  \Omega_{k=3}(\widehat{1},\widehat{i-1},\widehat{i},\widehat{j},\widehat{j+1}) \:,\label{NMHV}
\end{align}
where any $\Omega_{k=2}$ with less than 3 arguments is defined to be 1, and these shifted momenta are
\begin{align*}
\tilde{\lambda}_{\hat{1}} &=\tilde{\lambda}_{1}+\sum_{a=2}^{i-1}\frac{%
\langle i\,  a\rangle }{\langle i\,  1\rangle }\tilde{\lambda}_{a}+\sum_{a=j+2}^{n}%
\frac{\langle a\,  j+1\rangle }{\langle 1\,  j+1\rangle }\tilde{\lambda}_{a}\:, \\
\tilde{\lambda}_{\hat{i}} &=\tilde{\lambda}_{i}+\sum_{a=2}^{i-1}\frac{%
\langle a\,  1\rangle }{\langle i\,  1\rangle }\tilde{\lambda}_{a} \:, \qquad
\tilde{\lambda}_{\widehat{i+1}} =\tilde{\lambda}_{i+1}+\sum_{a=i+2}^{j-1}%
\frac{\langle j\,  a\rangle }{\langle j\,  i+1\rangle }\tilde{\lambda}_{a} \:,\\
\tilde{\lambda}_{\hat{j}} &=\tilde{\lambda}_{j}+\sum_{a=m+1}^{n-1}%
\frac{\langle a\,  i+1\rangle }{\langle j\,  i+1\rangle }\tilde{\lambda}_{a} \:, \qquad
\tilde{\lambda}_{\widehat{j+1}} =\tilde{\lambda}_{j+1}+\sum_{a=m+1}^{n-1}%
\frac{\langle 1\,  a\rangle }{\langle 1\,  j+1\rangle }\tilde{\lambda}_{a} \:.
\end{align*}
Here we briefly comment on the difference between our result and the momentum twistor form for NMHV case. Of course, their BCFW representations are very similar, but they are very different $d\log$ forms. The momentum twistor form for N$^{k{-}2}$MHV has dimension $4(k{-}2)$, which is independent of the number of particles $n$, and for NMHV it has a very nice interpretation as the canonical cyclic polytope; while our form always has degree $(2n{-}4)$ (independent of $k$). As will be discussed later, even for NMHV the geometric interpretation of our differential forms are still not clear. One reason is that these forms also encode the information of the MHV prefactor, which has been striped off in the momentum twistor forms (more precisely, the latter is the form for the (super) Wilson loop). Of course the two can be connected by {\it e.g.} first inserting back the MHV prefactor for the super Wilson loop, and translating to non-chiral superspace, but we remark here that it is still an open question how to directly connect the two differential forms.

For a generic $n$ and $k$, the order of adding inverse-soft factor becomes important. For $k=2$, our forms are independent with the way of adding inverse-soft factor since they are independent with the triangulation of the $n$-gon. For $k=3$, we must begin with a black pentagon, then add inverse-soft factor in three edges of this pentagon as shown in (\ref{NMHV}), however, each white or black region is triangulation-independent, and the shift momenta are also independent with the triangulation of these white regions. We leave the systematic study of our form for any $n$ and $k$ to the future. As an illustrative example, in the appendix we give IS-decomposition for 8-pt N$^{2}$MHV case, which immediately gives the form.

\section{Differential forms for loop integrands %in N=4 SYM
}  \label{sec4}

It has been well established that any Grassmannian cell and its super-function is determined by the associated on-shell diagram~\cite{ArkaniHamed:2012nw}. All mathematical characterizations and properties of on-shell diagrams carry over to our new picture of differential form directly, except one transforms to non-chiral superspace and takes $\eta \to d\l$ and $\tilde\eta \to d\lb$. For example, there are two common elementary operations for on-shell diagrams, one is \emph{direct product} which simply puts two on-shell diagrams together, and the other is \emph{projection} which just identifies two external legs. These two operations in our picture of differential form are simply:
\begin{align}
    &{\text{direct product :}}\qquad \mathcal{F}^{n_{L}+n_{R},k_{L}+k_{R}}\delta^{4}(P) = \mathcal{F}^{n_{L},k_{L}}_{L}\delta^{4}(P_{L}) \wedge \mathcal{F}_{R}^{n_{R},k_{R}}\delta^{4}(P_{R}) \:, \label{dir_pro} \\
    &{\text{projection :}}\qquad\qquad  \widehat{\mathcal{F}}^{n,k} = \int \frac{d^{2}\lambda_{I}d^{2}\tilde{\lambda}_{I}}{\mathrm{GL}(1)} \mathcal{F}^{n+2,k+1}(1,\cdots,n,I,-I)\vert_{d^{2}\lambda_{I}d^{2}\tilde{\lambda}_{I}} \:,
\end{align}
note that here we used the dressed form $\mathcal{F}$ instead of $\Omega$, it is still need to strip the overall factor $(dq)^{4}$ (or equivalently, $(d\tilde{q})^{4}$) to obtain the final result, and $\mathcal{F}\vert_{d^{2}\lambda_{I}d^{2}\tilde{\lambda}_{I}}$ means taking the coefficient of $d^{2}\lambda_{I} d^{2}\tilde{\lambda}_{I}$ in the dressed form $\mathcal{F}$.

In principle, these two operations, direct products and projections, together with two 3-pt differential forms are enough to build the corresponding form for any on-shell function. There are some specific combinations of these two operations simplify,
one is IS-construction mentioned above which is already enough to yield differential forms for all tree amplitudes. Generically, the on-shell functions can either have singular kinematic support for $d<2n{-}4$, or they can become differential forms that depend on some internal auxiliary variables as well for $d>2n{-}4$. There is a operation, that is so-called \emph{BCFW-bridge}~\cite{ArkaniHamed:2012nw}, yielding such differential forms with internal auxiliary variables.

In the context of on-shell diagram, adding a BCFW-bridge means the following operation:
\begin{equation*}
\raisebox{-65pt}{\includegraphics[scale=0.5]{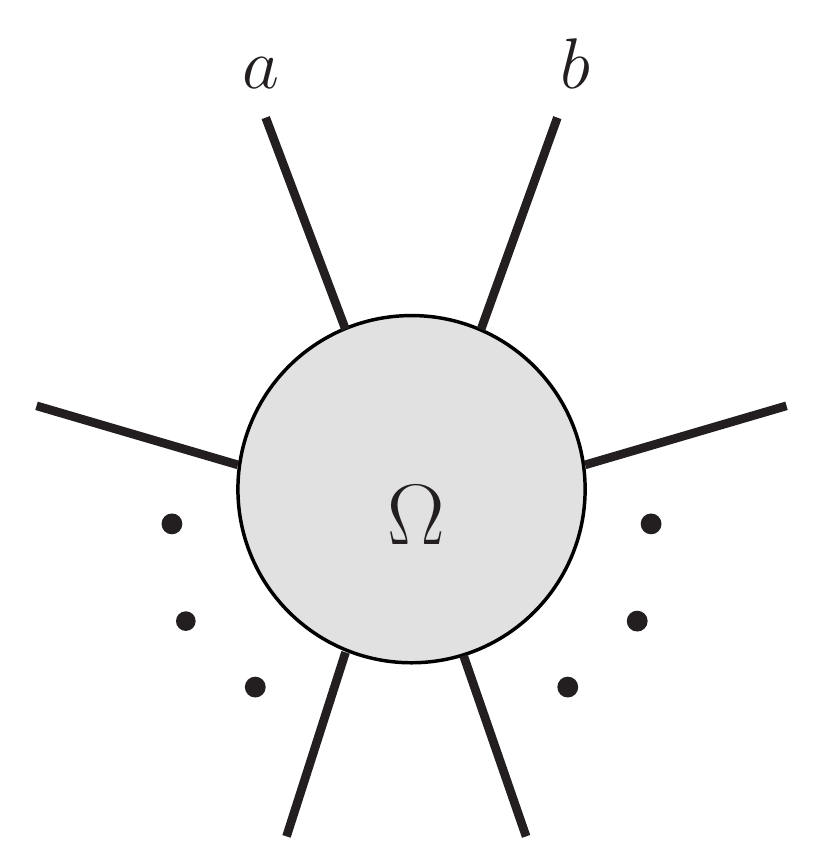}} \raisebox{-10pt}{\scalebox{2}[2]{$\implies$}}
\raisebox{-65pt}{\includegraphics[scale=0.5]{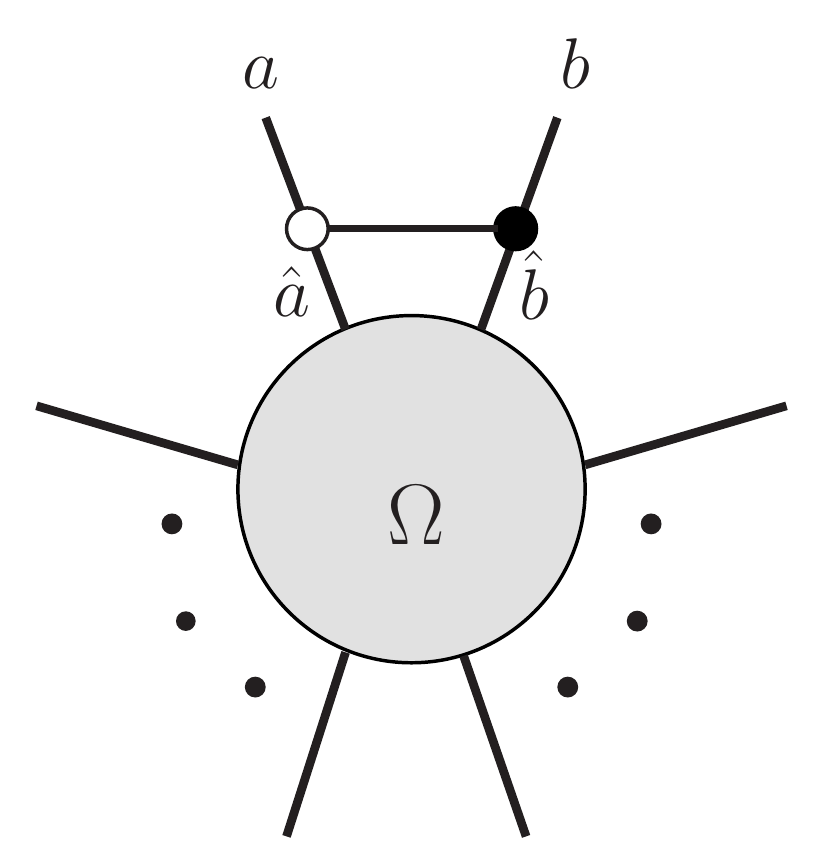}}
\end{equation*}
The net result of this operation just add a $d\log\alpha$ into the measure and shift the external date $a$ and $b$ by $\tilde{\lambda}_{a}\to \tilde{\lambda}_{\widehat{a}}= \tilde{\lambda}_{a}-\alpha \tilde{\lambda}_{b}$ and $\lambda_{b}\to\lambda_{\widehat{b}} =\lambda_{b}+\alpha\lambda_{a}$. Based on our push-forward definition (\ref{formGra1}), under this operation,
\begin{equation}
\Omega(1,\cdots,a,b,\cdots,n)\to d\log\alpha\: \Omega(1,\cdots,\widehat{a},\widehat{b},\cdots,n)
\end{equation}

\begin{comment}
The same argument applies to general $d$-dimensional cell in the Grassmannian $G_{+}(k,n)$, and the push-forward always gives the form for the corresponding on-shell functions, after removing $\delta^{4|4}(P|\tilde Q)$. Generically the on-shell functions can either have singular kinematic support, for $d<2n{-}4$, or for $d>2n{-}4$, they can become differential forms that depend on some internal, auxiliary variables as well~\cite{}.
\end{comment}

In the context of on-shell functions, such on-shell functions with auxiliary variables are $d\log$ forms dressed with super-functions that are known as leading singularities~\cite{ArkaniHamed:2010gh} (similar to BCFW terms). Here, an important point is that we no longer need to distinguish between such super-functions and the (internal) differential forms. By considering encoding helicity information of ${\cal N}=4$ SYM as a form on $\Gamma_n$, we have a unified picture: any Grassmannian cell evaluates to a $d\log$ form (possibly with singular kinematic supports) as the pushforward of its canonical form:
\be
f_{n,k}^{(\gamma)} \implies \Omega_{n,k}^{(\gamma)}=\omega_{n,k}^{(\gamma)}(\Lambda, \tilde\Lambda; \alpha)=d\log x_1 \wedge \cdots \wedge d\log x_d\ ,
\ee
which depends on external variables on $\Gamma_n$ and internal variables, collectively called $\alpha$.

\begin{comment}
 %However, it is still non-trivial and interesting to see how do we glue on-shell three-point amplitudes together in the form picture.
For one-shell super-functions in chiral superspace, one needs to do Grassmann integral over $d^4\eta$ for the sum over supermultiplet, and similarly over $d^2\eta\,d^2\tilde\eta$ in non-chiral superspace. This is equivalent to extracting the ``top component鈥? or the part with $d^2\eta\,d^2\tilde\eta$, of the super-function containing an internal edge for which we do the state sum, and the same is true for the differential form. Explicitly, when we have an internal edge with on-shell state $(\l, \lb)$, the state sum amounts to extracting  the coefficient of $(d\l)^2\,(d\lb)^2$ for the differential form.
With this new ingredient, one can compute the form for any on-shell diagrams by gluing together basic three-point amplitudes.

By consecutively attaching either the $k$-preserving or $k$-increasing inverse soft factors, it is obvious that one can obtain the MHV differential form, \eqref{MHV}. Another example, is by attaching one $k$-preserving and one $k$-increasing factors to the four-point form \eqref{4pt}, one obtains the 6pt NMHV form \eqref{6ptNMHV} and \eqref{6ptNMHV2}.
\end{comment}

Beyond tree amplitudes, we are mostly interested in forms that correspond to $L$-loop integrands in ${\cal N}=4$ SYM, which are generally $d=2n{-}4{+}4L$ forms. The $d$ canonical variables $x_1, \ldots, x_d$ can be solved in terms of external and loop momenta, and the push-forward simply gives $\wedge_{i=1}^d d\log x_i (\Lambda, \tilde\Lambda, \ell_1, \ldots, \ell_L)$. Note that we do not have to restrict ourselves to planar case, although for that we have the additional advantage that loop variables can be defined in an unambiguous way. In the old way of thinking about the loop integrand in ${\cal N}=4$ SYM, we have $4L$ $d\log$ forms which encode the loop integral measure, dressed with leading singularities as super-functions of external data. Now the picture unifies trees and loops: we have a combined $2n{-}4{+}4L$ $d\log$ forms with arguments depending on both external and loop variables. For example, since the one loop 4-pt amplitude can be obtained by adding 4 BCFW bridges between 4 external legs, the one-loop four-point form can be simply written as
\begin{align}
\Omega_{4,2}^{\text{1-loop}} &= d\log \alpha_{1}\,d\log \alpha_{2}\, d\log \alpha_{3} \, d\log \alpha_{4}\,
d\log \frac{\lan 1\,\widehat{2}\ran}{\lan 1\,3\ran}\, d\log \frac{\lan \widehat{2}\,3\ran}{\lan 1\,3\ran}\, d\log \frac{\lan 3\,\widehat{4}\ran}{\lan 1\,3\ran} \, d\log \frac{\lan \widehat{4}\,1\ran}{\lan 1\,3\ran} \nonumber \\
&= d\log \left(\frac{\alpha_{1}\langle 3\,4\rangle}{\langle 3\,4\rangle+\alpha_{1}\langle 3\,1\rangle}\right)
d\log \left(\frac{\alpha_{2}\langle 2\,3\rangle}{\langle 2\,3\rangle+\alpha_{2}\langle 1\,3\rangle}\right)\nonumber \\ &\quad d\log \left(\frac{\alpha_{3}\langle 1\,2\rangle}{\langle 1\,2\rangle+\alpha_{3}\langle 1\,3\rangle}\right)  d\log \left(\frac{\alpha_{4}\langle 4\,1\rangle}{\langle 4\,1\rangle+\alpha_{4}\langle 3\,1\rangle}\right)
\wedge \Omega_{4,2}^{\text{tree}}  \:,
\end{align}
here we omitted almost wedge product symbols for saving space, $\Omega_{4,2}^{\text{tree}}$ in the last row is the 4-pt tree differential form (\ref{4pt}), and the shifted external momenta are
\begin{equation*}
\lambda_{\widehat{2}} = \lambda_{2} + \alpha_{2}\lambda_{1} +\alpha_{3}\lambda_{3}  \qquad
\lambda_{\widehat{4}} = \lambda_{4} + \alpha_{1}\lambda_{1} +\alpha_{4}\lambda_{3}
\end{equation*}
The $d\log$ from besides the 4-pt tree form in this expression is exactly the 1-loop 4-pt amplitude integrand. In terms of the well-known four $d\log$ form for the loop,  the whole expression can also be written as
\begin{align}\label{1loop4pt}
\Omega^{\rm 1-loop}_{4,2}=d \log  \frac{\ell^2}{(\ell{-}\ell_*)^2} \wedge d \log  \frac{(\ell{+}p_1)^2}{(\ell{-}\ell_*)^2} \wedge d \log  \frac{(\ell{+}p_1{+}p_2)^2}{(\ell{-}\ell_*)^2} \wedge d \log  \frac{(\ell{-}p_4)^2}{(\ell{-}\ell_*)^2} \wedge \Omega^{\rm tree}_{4,2}\,,
\end{align}
here $\ell_{\ast}$ is a solution such that all the four propagators become on shell when plugging $\ell=\ell_{\ast}$. An interesting new features is that now the $d$ can hit external data as well, so the form contains not only $d^4\ell $ part but also $d^3 \ell, d^2\ell, d\,\ell$ etc. although the latter obviously integrate to zero with the usual ${\mathbb R}^{4}$ contour.

The Grassmannian/OSD picture makes it manifest that the $2n{-}4{+}4L$-form for $L$-loop integrand of planar ${\cal N}=4$ SYM is a sum of $d\log$'s. Obviously there are other representations as well, for example, one can translate the local representation~\cite{ArkaniHamed:2010gh}, {\it i.e.} a sum of pure $d\log$ integrals with local poles multiplied by leading singularities, into differential forms. As the first non-trivial example, we record the 1-loop 5-pt MHV form
\begin{align}
    \Omega_{5,2}^{\text{1-loop}} &= \biggl(d\log \frac{(\ell{-}p_{1})^{2}}{\ell^{2}}\wedge d\log\frac{(\ell{-}p_{1}{-}p_{2})^{2}}{\ell^{2}}\wedge d\log \frac{(\ell{-}p_{1}{-}p_{2})^{2}}{(\ell{-}\ell_{\ast}^{(1)})^{2}}\wedge d\log \frac{(\ell{+}p_{4}{+}p_{5})^{2}}{(\ell{-}\ell_{\ast}^{(1)})^{2}}\nonumber \\
    &\quad + d\log \frac{(\ell{+}p_{2}{+}p_{3})^{2}}{(\ell{-}p_{4}{-}p_{5})^{2}}\wedge d\log \frac{%
(\ell{+}p_{3})^{2}}{(\ell{-}p_{4}{-}p_{5})^{2}}\wedge d\log \frac{\ell^{2}}{(\ell{-}\ell_{\ast}^{(2)})^{2}}\wedge d\log \frac{(\ell{-}p_{4})^{2}}{(\ell{-}\ell_{\ast}^{(2)})^{2}} \nonumber\\
&\quad + d\log \frac{\ell^{2}}{(\ell{+}p_{1}{+}p_{2})^{2}}\wedge d\log \frac{%
(\ell{-}p_{3})^{2}}{(\ell{+}p_{1}{+}p_{2})^{2}}\wedge d\log \frac{(\ell{-}p_{3})^{2}}{(\ell{-}\ell_{\ast}^{(3)})^{2}}\wedge d\log \frac{(\ell{-}p_{3}{-}p_{4})^{2}}{(\ell{-}\ell_{\ast}^{(3)})^{2}}\biggr) \nonumber\\
&\quad \wedge \Omega^{\text{tree}}_{5,2} \label{1loop5pt}
\end{align}
which can be obtained from the one loop local form by turning momentum twistor variables into spinor variables~\cite{ArkaniHamed:2012nw} and corresponds to the following three integral,
\begin{equation*}
\raisebox{-65pt}{\includegraphics[scale=0.3]{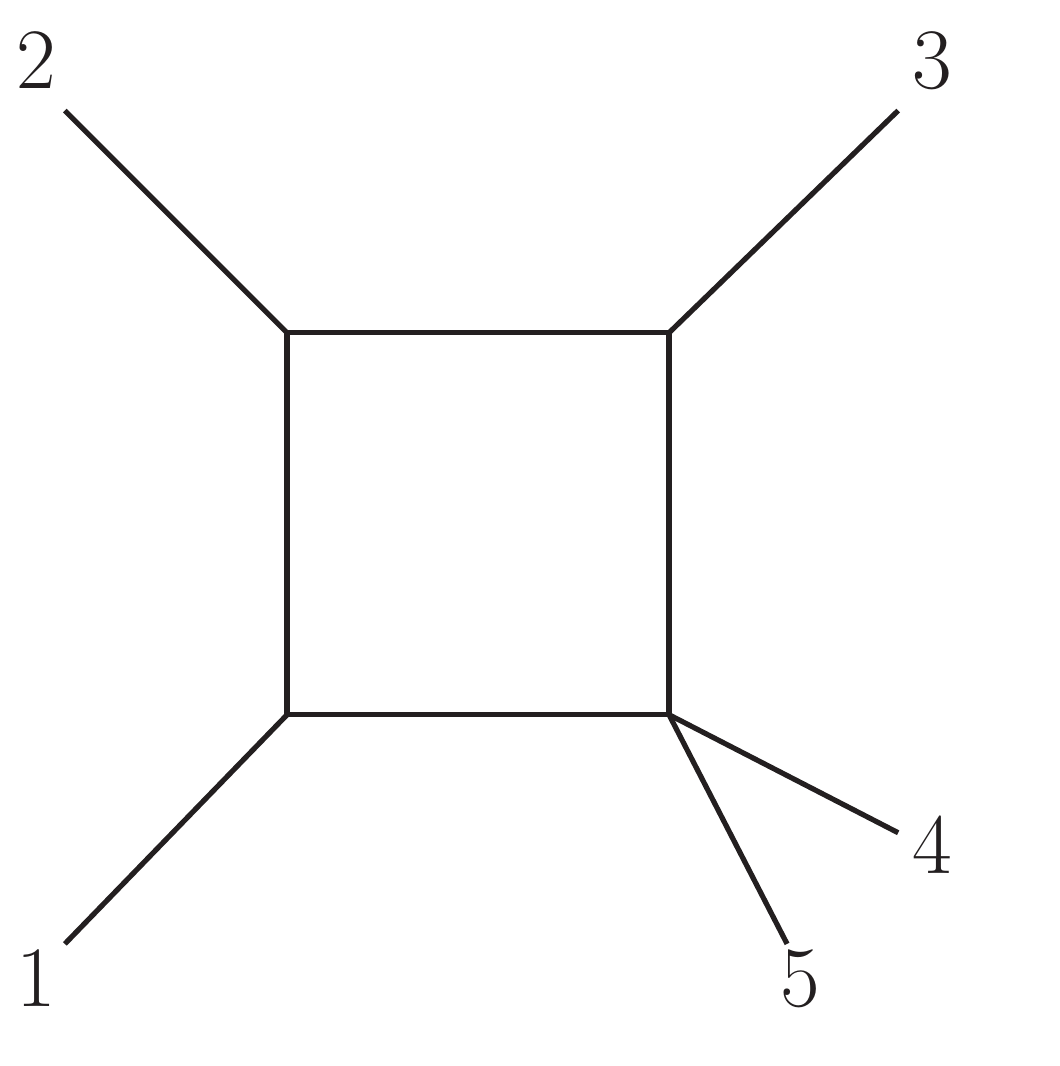}} \qquad
\raisebox{-65pt}{\includegraphics[scale=0.3]{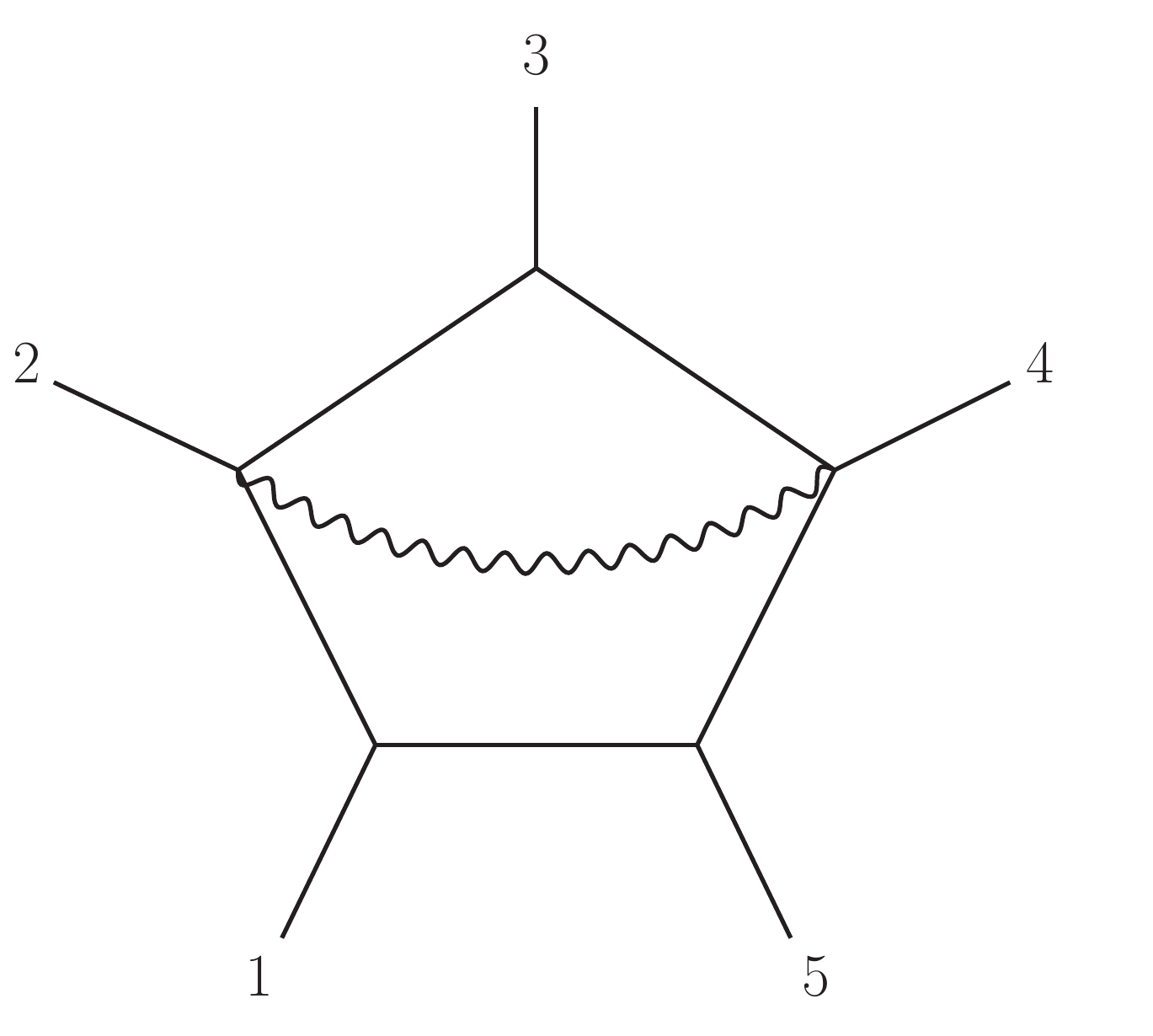}} \qquad
\raisebox{-65pt}{\includegraphics[scale=0.3]{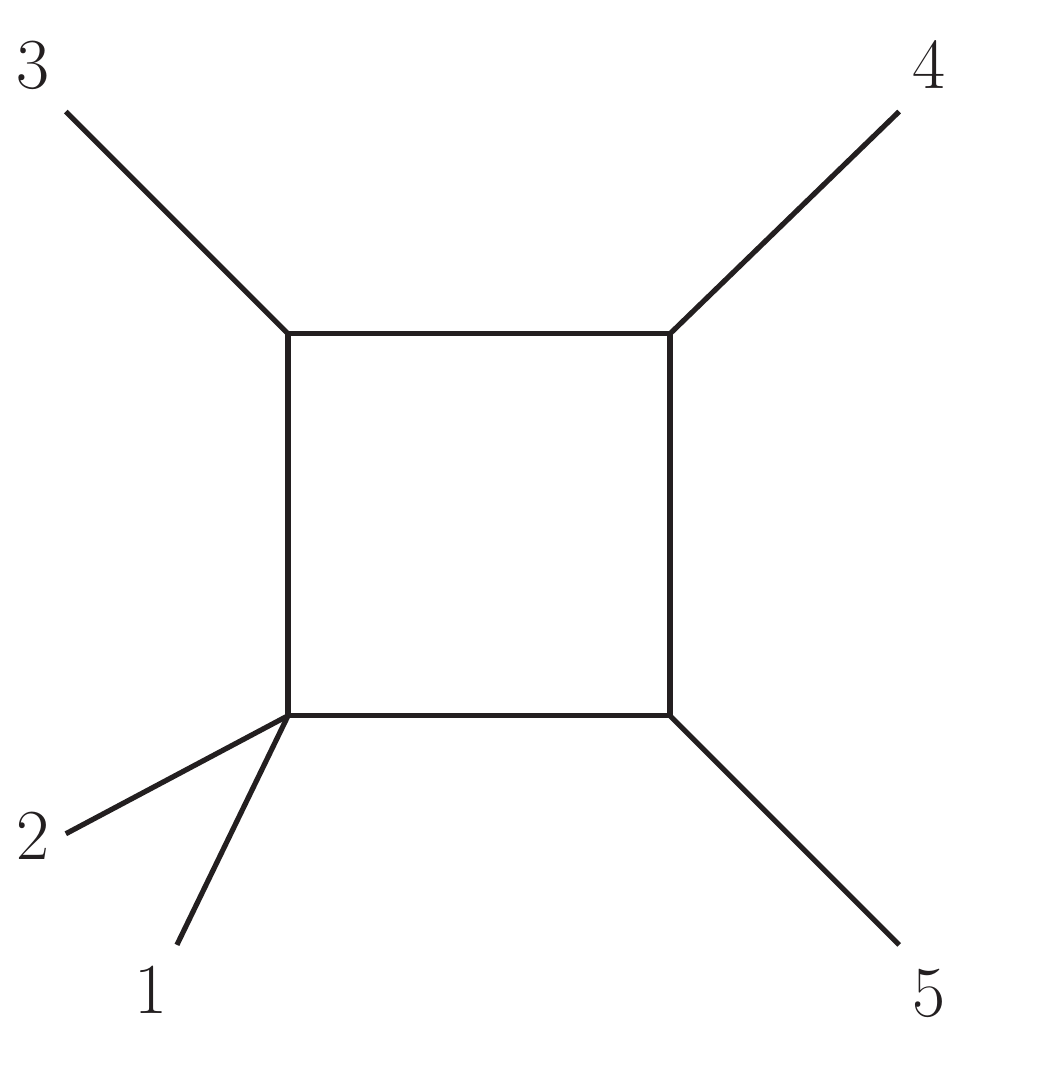}}
\end{equation*}
where
\begin{equation*}
    \ell_{\ast}^{(1)}=\biggl(\lambda_{1}+\frac{[32]}{[31]}\lambda_{2}\biggr)\tilde{\lambda}_{1}\:,\qquad
    \ell_{\ast}^{(2)}=\biggl(\lambda_{4}+\frac{[15]}{[14]}\lambda_{5}\biggr)\tilde{\lambda}_{4} \:,\qquad
    \ell_{\ast}^{(3)}=\lambda_{3}\biggl(\tilde{\lambda}_{3}+\frac{\langle54\rangle}{\langle53\rangle}\lambda_{4}\biggr)
\end{equation*}
are solutions such that some propagators on-shell, more precisely, $\ell_{\ast}^{(1)}$ is the solution of $\ell^{2}=(\ell{-}p_{1})^{2}=(\ell-p_{1}-p_{2})^{2}=(\ell+p_{4}+p_{5})^{2}=0$~\cite{Elvang:2015rqa}, while $\ell_{\ast}^{(2)}$ is the solution of $\ell^{2}=(\ell{-}p_{4})=(\ell{-}p_{4}-p_{5})^{2}=(\ell+p_{2}+p_{3})^{2}=0$, and similarly for $\ell_{\ast}^{(3)}$.

In general, an important question is how to characterize and determine the form for the planar integrand for ${\cal N}=4$ SYM. %In the next section we will see that at least for the tree form, there is an intrinsic geometric definition of the form which is completely parallel to the (tree) amplituhedron in momentum-twitstor space.
Here we show that the form can be completely characterized by residues on all its co-dimension one boundaries, which are factorization and single-cut poles. Of course the statement is parallel to that for usual super-amplitudes, but given the new way of summing over supermultiplet for the form, conceptually the form language makes the "hiding-particle" picture for loops more transparent. There are two types of simple poles for the loop integrand, factorization poles with $(\sum_{i\in L} p_i)^2=(\sum_{i\in R} p_i)^2\to 0$ and single cuts with $\ell^2\to 0$ for some loop variable $\ell$ (or more generally $(\ell+\sum_i p_i)^2$). Note that since we are taking the $d$ of external data as well, now factorization pole can also be seen as single cuts where we cut only external data. The key difference of these two types of cuts is that the former cut the amplitude into two parts $\Omega_L, \Omega_R$, connected by an on-shell internal leg, while the latter cut a loop and make it a lower-loop amplitude $\Omega^{(L{-}1)}_{n{+}2}$ with two additional legs under the forward limit.

Let's denote the two on-shell particles, as in factorization or forward limit, as $(\lambda, \tilde\lambda)$ and $ (\lambda, -\tilde\lambda)$, and recall that we need to extract the top component, {\it i.e.} taking the $(d\lambda)^2 (d\tilde\lambda)^2$ part, which receives contributions from $(d\lambda)^2 \times (-d\tilde\lambda)^2, d\lambda_\alpha \times d\lambda^\alpha (-d\tilde\lambda)^2, \ldots, (d\tilde\lambda)^2 \times (d\lambda)^2$. From a simple counting we know that we must take the form itself with the coefficient: for factorization, this is because we have a $2 (n_L+1)-4+4 l_L$ form and $2(n_R+1)-4+ 4 l_R$ form, which precisely gives $2(n_L+n_R)-4+4(l_L+l_R)=2n-4+4L$ form as needed; for forward limit, we have a $2(n+2)-4+4(L{-}1)=2n+4L-4$ form, which is also what we need. This is quite interesting since it demonstrates the hiding-particle picture vividly: the top form $(d\lambda)^2 (d\tilde\lambda)^2$ of the forward-limit particles, directly translate into the loop integral measure in the single cut limit, which can be written as $d^4 \ell/{\rm vol~ GL}(1)$. There is no need to do ${\rm GL}(2)$ integral as in the case of momentum twistor space, and we see directly that loops just come from top component of the two hidden particles! To summarize, we have
\begin{align}\label{singularity}
&\operatorname{Res}_{p^2=0}\Omega^{(L)}_{n,k}=\sum_{\substack{l_L+l_R=L\\k_L+k_R=k}}~\Omega^{(l_L)}_{n_L, k_L}(L, \{\lambda, \tilde\lambda\})~\Omega^{(l_R)}_{n_R, k_R} (\{\lambda, -\tilde\lambda\}, R)|_{(d\lambda)^2 (d\tilde\lambda)^2}\,,\quad p=\sum_{i \in L} p_i=\lambda\,\tilde\lambda\,,\nonumber\\
&\operatorname{Res}_{\ell^2=0}\Omega^{(L)}_{n,k}=\Omega^{(L{-}1)}_{n{+}2,k{+}1} (1,\ldots, n, \{\lambda, \tilde\lambda\}, \{\lambda, -\tilde\lambda\})|_{(d\lambda)^2 (d\tilde\lambda)^2}\,,\quad \ell=\lambda\,\tilde\lambda\,,
\end{align}
where we have extracted the $(d\lambda)^2 (d\tilde\lambda)^2$ part in both cases. Thus we have specified residues of the form on all simple poles, which is sufficient information to determine the form of all-loop planar integrand. One way to achieve this is to use BCFW recursion, which is of course parallel to that for all-loop superamplitude in planar ${\cal N}=4$ SYM.

\section{Towards tree amplituhedron in momentum space}   \label{sec5}
So far our discussions for forms in ${\cal N}=4$ SYM are more or less based on  Grassmannian/on-shell diagram picture, thus it is important to ask if there is a more intrinsic definition of the full amplitude form (not as a sum over Grassmannian cells/on-shell diagrams)? This is already quite challenging for tree amplitudes, since the ``amplituhedron in momentum space", whose canonical form gives the amplitude form, is yet to be found. On the other hand, there is a interesting formula for tree amplitudes in momentum space that is still missing in momentum twistor space, namely the RSV ``connected formula" from Witten's twistor string theory. We will see that it provides a pushforward formula for the tree form $\Omega^{\rm tree}_{n,k}$ as a single object, which we expect to serve as an important guide for understanding the underlying positive geometry.

As already mentioned in the introduction, the formula expresses tree-level super-amplitude as an integral over $G_+(2,n)$ (the moduli space of twistor-string worldsheet), which is localized by scattering equations in four dimensions. These scattering equations can be written as the constraints for Grassmannian using the Veronese map from $G(2,n)$ to $G(k,n)$: ${\bf C}(t, \sigma)_{\mu, a}=t_a \sigma_a^{\mu{-}1}$ for $a=1, 2, \ldots,n$ and $\mu=1, 2, \ldots, k$. Since we use the parity-invariant form, we can simply take the orthogonal complement G$(n{-}k,n)$ as ${\bf C}^{\perp}(\tilde t, \sigma)_{\tilde\mu, a}=\tilde t_a \sigma_a^{\tilde\mu{-}1}$ for $\tilde\mu=1, 2, \ldots, n{-}k$. Here $t$'s and $\tilde t$'s satisfy $t_a \tilde t_a=\prod_{b\neq a} (\sigma_a-\sigma_b)^{-1}$ (which implies $ {\bf C}_{\mu}\cdot {\bf C}^{\perp}_{\tilde\mu}=0_{k\times (n{-}k)}$ as expected).

In this parity-invariant form, the $(2n{-}4)$ canonical form of $G_+(2,n)$ reads~\cite{Cachazo:2013zc}
\be
\omega_n:=\frac 1 {{\rm vol\,GL}(2)}~\prod_{a=1}^n \frac{d\sigma_a~d t_a~d \tilde t_a}{(\sigma_a-\sigma_{a{+}1})}\, \delta\biggl(t_a \tilde t_a-\prod_{b\neq a}\frac 1 {(\sigma_a-\sigma_b)}\biggr).
\ee
where it is easy to see that we can recover the standard $G_+(2,n)$ form after integrating over $d \tilde t_a$'s (or $d t_a$'s) against the delta functions. Now the RSV formula is an integral over $\omega_n$ with delta functions similar to those in \eqref{onshellfunc}:
\begin{align}\label{RSV}
{\cal A}^{\rm tree}_{n,k}=\int \omega_n~\prod_{\mu=1}^k \delta^{(2|2)}\Bigl({\bf C}_{\mu} (t, \sigma)\cdot(\tilde\lambda |\tilde\eta)\Bigr)~\prod_{\tilde\mu=1}^{n{-}k} \delta^{(2|2)} \Bigl({\bf C}_{\tilde\mu}^{\perp}(\tilde t, \sigma)\cdot (\lambda|\eta)\Bigr)\,,
\end{align}
from which we can remove the overall delta functions for (super-)momentum conservations ${\cal A}^{\rm tree}_{n,k}:=\delta^{4|4}(P|\tilde Q)~A^{\rm tree}_{n,k}$, and again by replacing $\eta (\tilde\eta)$ by $d\lambda (d \tilde\lambda)$, we arrive at a formula for the form (here $\mu'$ denotes $k{-}2$ rows of ${\bf C}$ after removing those for overall delta functions)
\begin{align}\label{formRSV1}
\Omega^{\rm tree}_{n,k}=\int \omega_n~\prod_{\mu'} \delta^2({\bf C}_{\mu'}\cdot \tilde \lambda)\,({\bf C}_{\mu'}\cdot d \tilde\lambda)^2~\prod_{\tilde\mu}%=1}^{n{-}k}
\delta^2 ({\bf C}_{\tilde\mu}^{\perp}\cdot \lambda)\,({\bf C}_{\tilde\mu}\cdot d \lambda)^2\,,
%\nonumber\\&=\sum_{\rm sol.} \frac 1 {J_{n,k}~\prod_{a=1}^n (\sigma_a-\sigma_{a{+}1}) t_a}~\prod_{\mu'=3}^k ({\bf C}_{\mu',a} (t, \sigma) d \tilde\lambda_a)^2~\prod_{\tilde\mu=1}^{n{-}k}({\bf C}_{\tilde\mu, a} (\tilde t, \sigma) d \lambda_a)^2 \,,
\end{align}
From here the step is exactly the same as that from \eqref{formGra1} to \eqref{formGra2} except for one difference: instead of solving linear equations for a Grassmannian cell, we solve polynomial scattering equations and sum over their solutions $\{\sigma, t, \tilde t\}(\Lambda, \tilde\Lambda)$. Recall that in $k$-sector, we have Eulerian number $E(n{-}3, k{-}2)$ solutions, for $k=2,\cdots, n{-}2$. We arrive at:
\begin{align}\label{formRSV2}
\Omega^{\rm tree}_{n,k}=\sum_{\rm sol.}^{E(n{-}3, k{-}2)} \omega_n (\Lambda, \tilde\Lambda)%|_{\rm sol.}
=\sum_{\rm sol.} \frac 1 {{\rm vol\,GL}(2)}\prod_{a=1}^n \frac{d\sigma_a~d t_a}{\sigma_{a,a{+}1}~t_a} = {\rm parity~conjugate}\,,
%\sum_{\rm sol.} \frac 1 {{\rm vol\,GL}(2)}\prod_{a=1}^n \frac{d\sigma_a~d \tilde t_a}{\sigma_{a,a{+}1}~\tilde t_a}\,,
\end{align}
where by parity conjugate we mean $\omega_n$ written in terms of $\sigma$'s and $\tilde t$'s (thus $t$'s are solved).

Though equivalent to eq.\eqref{RSV}, eq.\eqref{formRSV2} itself is a remarkably pushforward formula from $\omega_n$ to $\Omega_{n,k}^{\rm tree}$. It strongly suggests the existence of a $(2n{-}4)$-dim postive geometry in $\Gamma_n$, which should be called the tree amplituhedorn in momentum space; the scattering equations, ${\bf C} (\sigma,t) \cdot \tilde\Lambda={\bf C}^{\perp} (\sigma, \tilde t) \cdot \Lambda=0$, in turn provide a map from $G_+(2,n)$ to it. The BCFW representation, or equivalently Grassmannian/OSD picture, provides certain triangulations of this object, just like in momentum twistor space, but the definition of the tree amplituhedron (and its canonical form) in $\Gamma_n$ is intrinsic and independent of any triangulation.

To define the tree amplituhedron in $\Gamma_n$, it is important to compare it with that in momentum twistor space. By analogy, it should be the intersection of a ``positive region" $\Gamma^+_{k,n}\subset \Gamma_n$, with a $(2n{-}4)$-dim subspace. While we have not been able to identify the subspace in general, here we present our conjecture for the ``positive region". Recall that a necessary condition for the push-forward is that for $\Lambda, \tilde\Lambda \in \Gamma^+_{k,n}$, the equations ${\bf C} (\sigma,t) \cdot \tilde\Lambda={\bf C}^{\perp} (\sigma, \tilde t) \cdot \Lambda=0$ have {\it exactly} one solutions in $G_{+}(2,n)$, which is defined as $\sigma_a<\sigma_{a{+}1}$ and either $t_a>0$ or $\tilde t_a>0$, for $a=1,2,\ldots,n$. Our proposal for $\Gamma^+_{k,n}$ consists of two conditions, which are the analog of those in the momentum twistor space~\cite{Arkani-Hamed:2017vfh}\footnote{The region can be translated from the region in momentum twistor space once we make a choice about where to put the infinity twistor. As pointed out to us by Arkani-Hamed, our choice amounts to putting the infinity twistor inside the one-loop amplituhedron or its parity conjugate.}:
\begin{itemize}
\item Positive kinematics: {\it i.e.} all planar poles $s_{i,i{+}1,\ldots, i{+}m}>0$.
\item Correct sign flips: let $N$ and $\tilde N$ to denote the sign flip in the list $\{ \lan 1\,2\ran, \lan 1\,3\ran, \ldots, \lan 1\,n\ran\}$, and $\{[1\,2], [1\,3], \ldots, [1\,n]\}$, respectively, here we require one of the two possibilities, $(N, \tilde N)=(k{-}2, k)$ or $(N,\tilde N)=(n{-}k, n{-}k{-}2)$.
\end{itemize}

For positive kinematics with $(N, \tilde N)=(k{-}2, k)$ we conjecture that there is exactly one solution to the scattering equations with $\sigma_a<\sigma_{a{+}1}$ and $t_a>0$, while for $(N,\tilde N)=(n{-}k, n{-}k{-}2)$ exactly one solution with $\sigma_a<\sigma_{a{+}1}$ and $\tilde t_a>0$. We do not have a proof for this, but it has been checked numerically for all $k$ sectors up to $n=10$. This is very strong evidence that the 4d scattering equations can be viewed as a one-to-one map from $G_+(2,n)$ to (certain $(2n{-}4)$-dim subspace of) this ``positive region".

Note that these equations are the refined version of the CHY scattering equations in four dimensions~\cite{Cachazo:2013iaa,He:2016vfi}: given 4d kinematics, the $(n{-}3)!$ solutions of scattering equations, %$\sum_{b\neq a} \frac{\langle a\, b\rangle~[a\,b]}{\sigma_{a,b}}=0$ for $a=1,2, \cdots, n$,
fall into $k=2,\cdots, n{-}2$ sectors which are exactly the solutions for $\sigma$'s in RSV equations, and for each solution we can get one solution for $t$'s (or $\tilde t$'s). Our observation here is thus a refinement of the observation in~\cite{Arkani-Hamed:2017mur} where it was found that for positive kinematics there is a unique solution with $\sigma_a<\sigma_{a{+}1}$. What is more here is that depending on what sign flips the spinors have, we can further identify the unique solution to a particular $k$ sector with all $t$'s (or $\tilde t$'s) being positive.

Last but not least, it is interesting to compare the complete tree form $\Omega^{{\cal N}=4, \rm tree}_n:=\sum_{k=2}^{n{-}2} \Omega^{\rm tree}_{n,k}$ with the $(n{-}3)$ canonical form of the associahedron for bi-adjoint $\phi^3$ amplitudes. The latter is given by the pushforward of the canonical form of ${\cal M}^+_{0,n}:=G_+(2,n)/GL(1)^{n{-}1}$, via summing over $(n{-}3)!$ solutions of the scattering equations~\cite{Arkani-Hamed:2017mur}:
\be
\Omega_{\phi^3}^{(n{-}3)}=\sum_{\rm sol}^{(n{-}3)!} \frac 1 {{\rm vol\,SL}(2)}\prod_{a=1}^n \frac{d\sigma_a}{\sigma_{a,a{+}1}}\,.
\ee
We see that, when summing over all solutions of 4d scattering equations, by dressing with $n{-}1$ $\frac{d t}{t}$'s or $\frac{d\tilde t}{\tilde t}$'s for each solution, it seems we can ``uplift" $\Omega_{\phi^3}^{(n{-}3)}$ to $\Omega^{(2n{-}4)}_{{\cal N}=4}$!

So far this is just an observation and we do not know if it implies any deeper connection between $\Omega_{\phi^3}$ and $\Omega_{{\cal N}=4}$. It does reflect a simple fact: the singularity structure of $\Omega_{{\cal N}=4}$ is given by those of $\Omega_{\phi^3}$, or planar cubic tree graphs, as well as soft singularities which correspond to either some $\lambda$ (equivalently $t$) or $\tilde\lambda$ ($\tilde t$) vanishes. This may provide another clue for finding the positive geometry underlying ${\cal N}=4$ tree form.

\section{Outlook}  \label{sec6}
In this paper, we have initiated the study of differential forms for scattering amplitudes in the space of spinor variables, and in particular the remarkable $d\log$ forms for ${\cal N}=4$ SYM. The main results can be summarized as follows
\begin{itemize}
\item It is natural to combine all helicity amplitudes of a theory into a differential form on $\Gamma_n$, which amounts to ``bosonizing" super-amplitudes for the supersymmetric case.
\item Tree amplitudes in ${\cal N}=4$ SYM yield a $d\log$ form, $\Omega^{\rm tree}_{n,k}$, which is manifest in the Grassmannian picture. We present explicit formula for MHV and NMHV forms. %A particularly convenient way for constructing it explicitly is via the inverse-soft construction, and we apply it to MHV and NMHV cases.
\item For any Grassmannian cell the form on an extended space can be obtained via pushforward, which in particular applies to the form for loop integrands in ${\cal N}=4$ SYM.
\item Our explorations strongly suggest an underlying ``amplituehdron in momentum space". %and we provide some hints for unraveling its geometry.
\end{itemize}

There are numerous open questions raised by these first steps, and here we discuss two directions concerning ${\cal N}=4$ SYM and more general theories.
\paragraph{Forms and amplituhedra in ${\cal N}=4$ SYM}
Obviously we have only scratched the surfaces of differential forms and geometries for amplitudes in ${\cal N}=4$ SYM. Already at tree level, the most pressing question and a major milestone in this direction, is to find the underlying amplituhedron geometry; if our conjecture is correct, all we need is to find a family of $(2n{-}4)$-dim subspaces. We emphasize that the positive geometries in $\Gamma_n$ must differ significantly from the tree amplituhedron in momentum twistor space, though both of them can be triangulated by BCFW terms. In this aspect, it would be highly desirable to look for some other triangulations (such as the analog of the ``local form" in~\cite{ArkaniHamed:2010gg}) or more intrinsic way of describing the geometry. For recent studies on triangulations and geometric structures of the usual amplituhedron, see~\cite{Bai:2014cna,Franco:2014csa,Bai:2015qoa,Ferro:2015grk,Karp:2016uax,Karp:2017ouj,Galashin:2018fri,Ferro:2018vpf}.

Beyond tree level, it is not clear at the moment how to get loop amplituhedron in a space extended by $L$ loop variables. Nevertheless, it is already of great interests to study differential forms for loop integrands. For example, we can compute various residues of the $L$-loop form: in addition to usual cuts which take residues with $d\ell$'s, one can also take residues with $d\lambda, d\tilde\lambda$'s {\it etc.}. Moreover, as already mentioned in~\cite{Arkani-Hamed:2017vfh}, the final (integrated) amplitudes now becomes some (pure) transcendental functions dressed by $d\log$ forms, which of course deserve further investigations.

A major advantage for ${\cal N}=4$ forms in spinor variables, as opposed to momentum twistor ones, is that we can study them beyond the planar limit.  Now we argue that the $L$-loop integrand in full (color-dressed) ${\cal N}=4$ SYM can also be written as a $2n{-}4{+}4L$ $d\log$ form, just as the planar case. This is based on two conjectures proposed recently: (1). those $4L$ loop-integral forms in non-planar ${\cal N}=4$ SYM only have logarithmic singularities~\cite{Arkani-Hamed:2014via} and (2). the leading singularities which dress them can be obtained from on-shell diagrams~\cite{ArkaniHamed:2012nw}, thus by translating into forms they can also be written as $2n{-}4$ $d\log$ forms. The upshot is the following conjecture. To all loop orders, one can find variables such that the form of full ${\cal N}=4$ SYM can be written as a sum of $d\log$'s:
\be
\Omega_{n,k}^{\rm L-loop}(\{\l,\lb, \ell\})=\sum_i~c_i~\left(\,\sum_{\alpha}~d\log x^{(i, \alpha)}_1 \wedge d \log x^{(i, \alpha)}_2 \wedge \cdots \wedge d \log x^{(i, \alpha)}_{2n{-}4{+}4L}\,\right)\,,
\ee
where we first sum over color structures $c_i$, and for each $i$ we sum over $d\log$ structures, $\alpha$'s, which can be {\it e.g.} (non-planar) on-shell diagrams or local $d\log$ integrals. Let's illustrate this by the simplest non-trivial example, the two-loop four-point integrand in full ${\cal N}=4$ SYM, whose form reads~\cite{Arkani-Hamed:2014via}
\be
\Omega^{\rm 2-loop}_{4, {\rm full}}=\sum_{\sigma\in S_4} \left( c^{(P)}_\sigma {\bf \Omega}^{(P)}_\sigma+ c^{(NP)}_\sigma {\bf \Omega}^{(NP)}_\sigma \right)\ ,\nonumber
\ee
where we sum over permutations $\sigma \in S_4$ and over planar and non-planar double boxes,
with $c$'s their color factors.  Here ${\bf \Omega}$'s are 12-forms obtained from the $d\log$ integrals in~\cite{Arkani-Hamed:2014via} and the \eqref{4pt} with different orderings: ${\bf \Omega}_{\sigma}^{(P)}=\Omega_{\sigma}^{\rm tree} \wedge \Omega^{(P)}_{\sigma}$ with $\Omega^{(P)}_{\sigma}$ the normalized double-box integral, and ${\bf \Omega}^{(NP)}_{(1234)}=\Omega_1 \wedge \Omega^{\rm tree}_{(1234)} \wedge \Omega_2^+ + \Omega^{\rm tree}_{(1243)} \wedge \Omega_2^-)$ where $\Omega_1$ is a $d\log$, one-loop sub-integral and $\Omega_2^{\pm}=\Omega_2^{\rm even} \pm \Omega_2^{\rm odd}$ with parity-even and parity-odd 4-forms.

In~\cite{Arkani-Hamed:2014bca}, a closed formula for MHV non-planar leading singularities has been obtained; by trivially taking the pushforward of the correspodning $G(2,n)$ cell, we record here the nice expression for their differential forms. Any such MHV on-shell diagram with $d=2n{-}4$ has $n{-}2$ black vertices (denote the set as $T$), where each vertex $\tau$ is connected (via white vertices) to three external legs $\tau\equiv\{\tau_1, \tau_2, \tau_3\}$, and the graph can be specified by $n{-}2$ such triplets $\tau$'s~\cite{Arkani-Hamed:2014bca}. The total form is nothing but the wedge product of the $(n{-}2)$ 2-forms, $\Omega_{3,2}(\tau_1, \tau_2, \tau_3)$, in \eqref{3pt2}:
%%$d\log \lan \tau_1\,\tau_2 \ran/\lan \tau 3\,\tau_1 \ran \wedge d\log \lan \tau 2\, \tau_3 \ran/\lan \tau_3\,\tau_1\ran$,
\be\label{MHVnp}
\Omega^{(T)}_{n,2}=\prod_{\tau \in T} d\log \frac{\lan \tau_1\,\tau_2 \ran}{\lan \tau_3\,\tau_1 \ran} \wedge d\log \frac{\lan \tau_2\, \tau_3 \ran}{\lan \tau_3\,\tau_1\ran}\ ,
\ee
which of course reduce to \eqref{MHV} for planar case. It would be interesting to generalize this to non-planar on-shell diagrams for higher $k$ ({\it c.f.}~\cite{Bourjaily:2016mnp}) and to connect to possible non-planar extension of amplituhedron~\cite{Bern:2015ple}.

 \paragraph{Forms in more general theories}

Our discussions have centered around ${\cal N}=4$ SYM, but one important direction is to study forms in more general theories. First of all, as we have outlined, one can write down forms for less or non-supersymmetric theories, which are no longer $d\log$ forms. In addition to logarithmic singularities, these forms have poles at infinity. To be concrete, let's even consider the $2n$-form that combines all MHV gluon amplitudes in pure Yang-Mills:
\be
\Omega^{\rm YM}_{n, {\rm MHV}}=\sum_{i,j} \frac{\langle i\,j\rangle^4 (d\tilde\lambda_i)^2 (d\tilde\lambda_j)^2 \prod_{k\neq i,j} (d\lambda_k)^2}{\langle 1\,2\rangle \langle 2\,3\rangle \cdots \langle n\,1\rangle}\,.
\ee
This is {\it not} a $d\log$ form, but it can be obtained by projecting the vanishing $(d q)^4 \Omega^{{\cal N}=4}_{n, {\rm MHV}}$ to gluonic part. It turns out that such a projection introduces poles at infinity, and already at tree level it would be interesting to systematically study them; we expect that such poles at infinity are responsible for boundary terms in BCFW recursion for gluon amplitudes~\cite{Feng:2009ei}, and in the Grassmannian/on-shell-diagram picture they are related to the Jacobian factors~\cite{ArkaniHamed:2012nw}. Beyond tree level, it would be of great importance to study forms for loop integrands in theories with less or no supersymmetries, since the additional poles must be related to UV property, which is absent in the ${\cal N}=4$ case.

Though we've restricted our discussions to gauge theories, it is trivial to write down differential forms for gravity amplitudes for $\mathcal{N} \leq 4$ (super) gravity {\it{e.g.}} from the double-copy of pure Yang-Mills amplitude (function) with forms for $\mathcal{N} \leq 4$ SYM. Of course we need many differential forms (one for each helicity amplitude from the pure Yang-Mills side) since $d\lambda, d\tilde\lambda$ cannot accommodate all helicity amplitudes for gravitons. Moreover, it is tempting to further extend the idea to massive theories and/or those in other dimensions. For example, using massive spinor-helicity variables developed in~\cite{Arkani-Hamed:2017jhn}, it is possible that one can combine massive amplitudes into differential forms, and in particular the form for ${\cal N}=4$ SYM on the Columb branch might again be the prototype of such massive forms. Going beyond four dimensions, it is also tempting to {\it e.g.} possible $d\log$ forms for ABJM amplitudes in $d=3$ and supersymmetric theories in $d=6$, where Grassmannian and related picture exists~\cite{Huang:2013owa,Huang:2014xza,Cachazo:2018hqa}.

\section*{Acknowledgement} We thank Nima Arkani-Hamed for first suggesting the problem, for numerous stimulating discussions and comments on the draft. We also thank Yuntao Bai, Yu-tin Huang, Jaroslav Trnka, Gongwang Yan and Yong Zhang for helpful discussions. S.H. thanks IAS Princeton for hospitality during a visit when the work started, and the organizers and participants of KITP conference "Scattering Amplitudes: from Gauge Theory to Gravity" (2017) where part of the work was first presented. S.H.'s research is supported in part by the Thousand Young Talents program, the Key Research Program of Frontier Sciences of CAS under Grant No. QYZDB-SSW-SYS014 and Peng Huanwu center under Grant No. 11747601.

\section{Appendix}

The IS decomposition of 8pt-$\mathrm{N}^{2}\mathrm{MHV}$:%
\begin{align*}
\{3,6,7,8,10,12,9,13\} &=(1,3,4,7)\otimes B(5,4,7)\otimes B(6,5,7)\otimes W(8,1,7)\otimes W(2,1,3) \\
\{5,7,6,8,10,12,9,11\} &=(2,4,6,7)\otimes W(3,2,4)\otimes B(5,4,6)\otimes B(1,2,7)\otimes W(8,1,7) \\
\{6,7,5,8,10,11,9,12\} &=(1,2,3,7)\otimes B(5,3,7)\otimes B(6,5,7)\otimes [W(4,3,5)\otimes W(8,1,7)] \\
\{5,6,7,10,11,8,9,12\} &=(1,2,3,6)\otimes B(4,3,6)\otimes B(5,4,6)\otimes W(7,8,1,6) \label{4.4}\\
\{4,6,7,10,8,11,9,13\} &=(1,4,5,7)\otimes [B(6,5,7)\otimes W(2,1,4)]\otimes B(3,2,4)\otimes W(8,1,7) \\
\{5,7,6,10,8,11,9,12\} &=(2,5,6,7)\otimes W(4,2,5)\otimes B(1,2,7)\otimes [W(3,2,4)\otimes B(8,1,7)] \\
\{3,5,7,8,10,9,12,14\} &=(3,4,5,6)\otimes B(1,3,6)\otimes W(8,1,6)\otimes W(2,1,3)\otimes B(7,8,6) \\
\{4,7,5,8,10,9,11,14\} &=(2,4,5,6)\otimes W(3,2,4)\otimes B(1,2,6)\otimes W(8,1,6)\otimes B(7,8,6) \\
\{4,5,7,10,8,9,11,14\} &=(1,2,3,5)\otimes B(4,3,5)\otimes W(8,1,5)\otimes W(6,5,8)\otimes B(7,6,8) \\
\{4,6,8,10,7,9,11,13\} &=(1,2,3,5)\otimes B(4,3,5)\otimes W(8,1,5)\otimes B(7,8,5)\otimes W(6,7,5) \\
\{3,4,7,8,9,10,13,14\} &=(1,4,5,8)\otimes [W(3,1,4)\otimes B(6,5,8)]\otimes [W(2,1,3)\otimes B(7,6,8)]
\\
\{3,6,8,9,7,10,12,13\} &=(1,3,4,8)\otimes B(5,4,8)\otimes B(7,8,5)\otimes [W(2,1,3)\otimes W(6,5,7)] \\
\{3,6,8,7,9,12,10,13\} &=(3,4,5,7)\otimes B(6,7,5)\otimes W(8,3,7)\otimes B(1,3,8)\otimes W(2,1,3) \\
\{3,5,8,7,9,10,12,14\} &=(3,4,5,6)\otimes W(8,3,6)\otimes B(1,3,8)\otimes [W(2,1,3)\otimes B(7,6,8)] \\
\{5,8,6,9,7,11,10,12\} &=(2,3,4,7)\otimes B(5,4,7)\otimes W(8,2,7)\otimes B(1,2,8)\otimes W(6,5,7) \\
\{4,8,6,9,7,10,11,13\} &=(2,4,5,8)\otimes W(3,2,4)\otimes B(1,2,8)\otimes B(7,1,8)\otimes W(6,5,8) \\
\{5,8,9,6,7,10,11,12\} &=(3,5,7,8)\otimes W(4,3,5)\otimes B(1,3,8)\otimes B(2,3,1)\otimes W(6,7,8) \\
\{5,8,6,7,9,12,10,11\} &=(2,6,7,8)\otimes W(4,2,6)\otimes W(3,2,4)\otimes [B(5,4,6)\otimes B(1,2,8)] \\
\{6,8,5,7,9,11,10,12\} &=(2,3,5,7)\otimes B(6,5,7)\otimes W(8,2,7)\otimes B(1,2,8)\otimes W(4,3,5) \\
\{4,8,5,7,9,10,11,14\} &=(2,5,6,8)\otimes W(4,2,5)\otimes W(3,2,4)\otimes B(1,2,8)\otimes B(7,6,8)
\end{align*}%
Notation: The BCFW scheme we used is $\{-2,2,0\}$. The left side of equations are the permutation of an OSD~\cite{Bourjaily:2012gy}, we begin with a colorless quadrangle (since the 4-pt amplitude can be MHV as well as $\overline{\mathrm{MHV}}$), and add some black or white triangle to make a 8-gon, the white triangle means k-preserving ISF, and black-triangle means k-increasing ISF, the order can not be changed unless they are in a square bracket [ ].

\bibliographystyle{utphys}
\bibliography{bib}

\providecommand{\href}[2]{#2}\begingroup\raggedright\begin{thebibliography}{10}

\bibitem{Arkani-Hamed:2017vfh}
N.~Arkani-Hamed, H.~Thomas, and J.~Trnka, ``{Unwinding the Amplituhedron in
  Binary},'' \href{http://dx.doi.org/10.1007/JHEP01(2018)016}{{\em JHEP}
  {\bfseries 01} (2018) 016},
\href{http://arxiv.org/abs/1704.05069}{{\ttfamily arXiv:1704.05069 [hep-th]}}.
%%CITATION = ARXIV:1704.05069;%%.

\bibitem{Arkani-Hamed:2013jha}
N.~Arkani-Hamed and J.~Trnka, ``{The Amplituhedron},''
  \href{http://dx.doi.org/10.1007/JHEP10(2014)030}{{\em JHEP} {\bfseries 10}
  (2014) 030},
\href{http://arxiv.org/abs/1312.2007}{{\ttfamily arXiv:1312.2007 [hep-th]}}.
%%CITATION = ARXIV:1312.2007;%%.

\bibitem{Hodges:2009hk}
A.~Hodges, ``{Eliminating spurious poles from gauge-theoretic amplitudes},''
  \href{http://dx.doi.org/10.1007/JHEP05(2013)135}{{\em JHEP} {\bfseries 05}
  (2013) 135},
\href{http://arxiv.org/abs/0905.1473}{{\ttfamily arXiv:0905.1473 [hep-th]}}.
%%CITATION = ARXIV:0905.1473;%%.

\bibitem{Arkani-Hamed:2017tmz}
N.~Arkani-Hamed, Y.~Bai, and T.~Lam, ``{Positive Geometries and Canonical
  Forms},'' \href{http://dx.doi.org/10.1007/JHEP11(2017)039}{{\em JHEP}
  {\bfseries 11} (2017) 039},
\href{http://arxiv.org/abs/1703.04541}{{\ttfamily arXiv:1703.04541 [hep-th]}}.
%%CITATION = ARXIV:1703.04541;%%.

\bibitem{Arkani-Hamed:2017mur}
N.~Arkani-Hamed, Y.~Bai, S.~He, and G.~Yan, ``{Scattering Forms and the
  Positive Geometry of Kinematics, Color and the Worldsheet},''
\href{http://arxiv.org/abs/1711.09102}{{\ttfamily arXiv:1711.09102 [hep-th]}}.
%%CITATION = ARXIV:1711.09102;%%.

\bibitem{Arkani-Hamed:2017fdk}
N.~Arkani-Hamed, P.~Benincasa, and A.~Postnikov, ``{Cosmological Polytopes and
  the Wavefunction of the Universe},''
\href{http://arxiv.org/abs/1709.02813}{{\ttfamily arXiv:1709.02813 [hep-th]}}.
%%CITATION = ARXIV:1709.02813;%%.

\bibitem{Gao:2017dek}
X.~Gao, S.~He, and Y.~Zhang, ``{Labelled tree graphs, Feynman diagrams and disk
  integrals},'' \href{http://dx.doi.org/10.1007/JHEP11(2017)144}{{\em JHEP}
  {\bfseries 11} (2017) 144},
\href{http://arxiv.org/abs/1708.08701}{{\ttfamily arXiv:1708.08701 [hep-th]}}.
%%CITATION = ARXIV:1708.08701;%%.

\bibitem{He:2018pue}
S.~He, G.~Yan, C.~Zhang, and Y.~Zhang, ``{Scattering Forms, Worldsheet Forms
  and Amplitudes from Subspaces},''
\href{http://arxiv.org/abs/1803.11302}{{\ttfamily arXiv:1803.11302 [hep-th]}}.
%%CITATION = ARXIV:1803.11302;%%.

\bibitem{Salvatori:2018aha}
G.~Salvatori, ``{1-loop Amplitudes from the Halohedron},''
\href{http://arxiv.org/abs/1806.01842}{{\ttfamily arXiv:1806.01842 [hep-th]}}.
%%CITATION = ARXIV:1806.01842;%%.

\bibitem{PhysRevLett.56.2459}
S.~J. Parke and T.~R. Taylor, ``Amplitude for $n$-gluon scattering,''
  \href{http://dx.doi.org/10.1103/PhysRevLett.56.2459}{{\em Phys. Rev. Lett.}
  {\bfseries 56} (Jun, 1986) 2459--2460}.
  \url{https://link.aps.org/doi/10.1103/PhysRevLett.56.2459}.

\bibitem{ArkaniHamed:2008gz}
N.~Arkani-Hamed, F.~Cachazo, and J.~Kaplan, ``{What is the Simplest Quantum
  Field Theory?},'' \href{http://dx.doi.org/10.1007/JHEP09(2010)016}{{\em JHEP}
  {\bfseries 09} (2010) 016},
\href{http://arxiv.org/abs/0808.1446}{{\ttfamily arXiv:0808.1446 [hep-th]}}.
%%CITATION = ARXIV:0808.1446;%%.

\bibitem{Huang:2011um}
Y.-t. Huang, ``{Non-Chiral S-Matrix of N=4 Super Yang-Mills},''
\href{http://arxiv.org/abs/1104.2021}{{\ttfamily arXiv:1104.2021 [hep-th]}}.
%%CITATION = ARXIV:1104.2021;%%.

\bibitem{Plefka:2014fta}
J.~Plefka, T.~Schuster, and V.~Verschinin, ``{From Six to Four and More:
  Massless and Massive Maximal Super Yang-Mills Amplitudes in 6d and 4d and
  their Hidden Symmetries},''
  \href{http://dx.doi.org/10.1007/JHEP01(2015)098}{{\em JHEP} {\bfseries 01}
  (2015) 098},
\href{http://arxiv.org/abs/1405.7248}{{\ttfamily arXiv:1405.7248 [hep-th]}}.
%%CITATION = ARXIV:1405.7248;%%.

\bibitem{ArkaniHamed:2010kv}
N.~Arkani-Hamed, J.~L. Bourjaily, F.~Cachazo, S.~Caron-Huot, and J.~Trnka,
  ``{The All-Loop Integrand For Scattering Amplitudes in Planar N=4 SYM},''
  \href{http://dx.doi.org/10.1007/JHEP01(2011)041}{{\em JHEP} {\bfseries 01}
  (2011) 041},
\href{http://arxiv.org/abs/1008.2958}{{\ttfamily arXiv:1008.2958 [hep-th]}}.
%%CITATION = ARXIV:1008.2958;%%.

\bibitem{Arkani-Hamed:2014bca}
N.~Arkani-Hamed, J.~L. Bourjaily, F.~Cachazo, A.~Postnikov, and J.~Trnka,
  ``{On-Shell Structures of MHV Amplitudes Beyond the Planar Limit},''
  \href{http://dx.doi.org/10.1007/JHEP06(2015)179}{{\em JHEP} {\bfseries 06}
  (2015) 179},
\href{http://arxiv.org/abs/1412.8475}{{\ttfamily arXiv:1412.8475 [hep-th]}}.
%%CITATION = ARXIV:1412.8475;%%.

\bibitem{Witten:2003nn}
E.~Witten, ``{Perturbative gauge theory as a string theory in twistor space},''
  \href{http://dx.doi.org/10.1007/s00220-004-1187-3}{{\em Commun. Math. Phys.}
  {\bfseries 252} (2004) 189--258},
\href{http://arxiv.org/abs/hep-th/0312171}{{\ttfamily arXiv:hep-th/0312171
  [hep-th]}}.
%%CITATION = HEP-TH/0312171;%%.

\bibitem{PhysRevD.70.026009}
R.~Roiban, M.~Spradlin, and A.~Volovich, ``Tree-level s matrix of yang-mills
  theory,'' \href{http://dx.doi.org/10.1103/PhysRevD.70.026009}{{\em Phys. Rev.
  D} {\bfseries 70} (Jul, 2004) 026009}.
  \url{https://link.aps.org/doi/10.1103/PhysRevD.70.026009}.

\bibitem{Elvang:2011fx}
H.~Elvang, Y.-t. Huang, and C.~Peng, ``{On-shell superamplitudes in N<4 SYM},''
  \href{http://dx.doi.org/10.1007/JHEP09(2011)031}{{\em JHEP} {\bfseries 09}
  (2011) 031},
\href{http://arxiv.org/abs/1102.4843}{{\ttfamily arXiv:1102.4843 [hep-th]}}.
%%CITATION = ARXIV:1102.4843;%%.

\bibitem{ArkaniHamed:2012nw}
N.~Arkani-Hamed, J.~L. Bourjaily, F.~Cachazo, A.~B. Goncharov, A.~Postnikov,
  and J.~Trnka, {\em {Grassmannian Geometry of Scattering Amplitudes}}.
\newblock Cambridge University Press, 2016.
\newblock \href{http://arxiv.org/abs/1212.5605}{{\ttfamily arXiv:1212.5605
  [hep-th]}}.
\newblock
\url{https://inspirehep.net/record/1208741/files/arXiv:1212.5605.pdf}.
\newblock
%%CITATION = ARXIV:1212.5605;%%.

\bibitem{ArkaniHamed:2010gh}
N.~Arkani-Hamed, J.~L. Bourjaily, F.~Cachazo, and J.~Trnka, ``{Local Integrals
  for Planar Scattering Amplitudes},''
  \href{http://dx.doi.org/10.1007/JHEP06(2012)125}{{\em JHEP} {\bfseries 06}
  (2012) 125},
\href{http://arxiv.org/abs/1012.6032}{{\ttfamily arXiv:1012.6032 [hep-th]}}.
%%CITATION = ARXIV:1012.6032;%%.

\bibitem{Elvang:2015rqa}
H.~Elvang and Y.-t. Huang, {\em {Scattering Amplitudes in Gauge Theory and
  Gravity}}.
\newblock Cambridge University Press, 2015.
\newblock
\url{http://www.cambridge.org/mw/academic/subjects/physics/theoretical-physics-and-mathematical-physics/scattering-amplitudes-gauge-theory-and-gravity?format=AR}.
\newblock
%%CITATION = INSPIRE-1384881;%%.

\bibitem{Cachazo:2013zc}
F.~Cachazo, ``{Resultants and Gravity Amplitudes},''
\href{http://arxiv.org/abs/1301.3970}{{\ttfamily arXiv:1301.3970 [hep-th]}}.
%%CITATION = ARXIV:1301.3970;%%.

\bibitem{Cachazo:2013iaa}
F.~Cachazo, S.~He, and E.~Y. Yuan, ``{Scattering in Three Dimensions from
  Rational Maps},'' \href{http://dx.doi.org/10.1007/JHEP10(2013)141}{{\em JHEP}
  {\bfseries 10} (2013) 141},
\href{http://arxiv.org/abs/1306.2962}{{\ttfamily arXiv:1306.2962 [hep-th]}}.
%%CITATION = ARXIV:1306.2962;%%.

\bibitem{He:2016vfi}
S.~He, Z.~Liu, and J.-B. Wu, ``{Scattering Equations, Twistor-string Formulas
  and Double-soft Limits in Four Dimensions},''
  \href{http://dx.doi.org/10.1007/JHEP07(2016)060}{{\em JHEP} {\bfseries 07}
  (2016) 060},
\href{http://arxiv.org/abs/1604.02834}{{\ttfamily arXiv:1604.02834 [hep-th]}}.
%%CITATION = ARXIV:1604.02834;%%.

\bibitem{ArkaniHamed:2010gg}
N.~Arkani-Hamed, J.~L. Bourjaily, F.~Cachazo, A.~Hodges, and J.~Trnka, ``{A
  Note on Polytopes for Scattering Amplitudes},''
  \href{http://dx.doi.org/10.1007/JHEP04(2012)081}{{\em JHEP} {\bfseries 04}
  (2012) 081},
\href{http://arxiv.org/abs/1012.6030}{{\ttfamily arXiv:1012.6030 [hep-th]}}.
%%CITATION = ARXIV:1012.6030;%%.

\bibitem{Bai:2014cna}
Y.~Bai and S.~He, ``{The Amplituhedron from Momentum Twistor Diagrams},''
  \href{http://dx.doi.org/10.1007/JHEP02(2015)065}{{\em JHEP} {\bfseries 02}
  (2015) 065},
\href{http://arxiv.org/abs/1408.2459}{{\ttfamily arXiv:1408.2459 [hep-th]}}.
%%CITATION = ARXIV:1408.2459;%%.

\bibitem{Franco:2014csa}
S.~Franco, D.~Galloni, A.~Mariotti, and J.~Trnka, ``{Anatomy of the
  Amplituhedron},'' \href{http://dx.doi.org/10.1007/JHEP03(2015)128}{{\em JHEP}
  {\bfseries 03} (2015) 128},
\href{http://arxiv.org/abs/1408.3410}{{\ttfamily arXiv:1408.3410 [hep-th]}}.
%%CITATION = ARXIV:1408.3410;%%.

\bibitem{Bai:2015qoa}
Y.~Bai, S.~He, and T.~Lam, ``{The Amplituhedron and the One-loop Grassmannian
  Measure},'' \href{http://dx.doi.org/10.1007/JHEP01(2016)112}{{\em JHEP}
  {\bfseries 01} (2016) 112},
\href{http://arxiv.org/abs/1510.03553}{{\ttfamily arXiv:1510.03553 [hep-th]}}.
%%CITATION = ARXIV:1510.03553;%%.

\bibitem{Ferro:2015grk}
L.~Ferro, T.~Lukowski, A.~Orta, and M.~Parisi, ``{Towards the Amplituhedron
  Volume},'' \href{http://dx.doi.org/10.1007/JHEP03(2016)014}{{\em JHEP}
  {\bfseries 03} (2016) 014},
\href{http://arxiv.org/abs/1512.04954}{{\ttfamily arXiv:1512.04954 [hep-th]}}.
%%CITATION = ARXIV:1512.04954;%%.

\bibitem{Karp:2016uax}
S.~N. Karp and L.~K. Williams, ``{The m=1 amplituhedron and cyclic hyperplane
  arrangements},''
\href{http://arxiv.org/abs/1608.08288}{{\ttfamily arXiv:1608.08288 [math.CO]}}.
%%CITATION = ARXIV:1608.08288;%%.

\bibitem{Karp:2017ouj}
S.~N. Karp, L.~K. Williams, and Y.~X. Zhang, ``{Decompositions of
  amplituhedra},''
\href{http://arxiv.org/abs/1708.09525}{{\ttfamily arXiv:1708.09525 [math.CO]}}.
%%CITATION = ARXIV:1708.09525;%%.

\bibitem{Galashin:2018fri}
P.~Galashin and T.~Lam, ``{Parity duality for the amplituhedron},''
\href{http://arxiv.org/abs/1805.00600}{{\ttfamily arXiv:1805.00600 [math.CO]}}.
%%CITATION = ARXIV:1805.00600;%%.

\bibitem{Ferro:2018vpf}
L.~Ferro, T.~Lukowski, and M.~Parisi, ``{Amplituhedron meets Jeffrey-Kirwan
  Residue},''
\href{http://arxiv.org/abs/1805.01301}{{\ttfamily arXiv:1805.01301 [hep-th]}}.
%%CITATION = ARXIV:1805.01301;%%.

\bibitem{Arkani-Hamed:2014via}
N.~Arkani-Hamed, J.~L. Bourjaily, F.~Cachazo, and J.~Trnka, ``{Singularity
  Structure of Maximally Supersymmetric Scattering Amplitudes},''
  \href{http://dx.doi.org/10.1103/PhysRevLett.113.261603}{{\em Phys. Rev.
  Lett.} {\bfseries 113} no.~26, (2014) 261603},
\href{http://arxiv.org/abs/1410.0354}{{\ttfamily arXiv:1410.0354 [hep-th]}}.
%%CITATION = ARXIV:1410.0354;%%.

\bibitem{Bourjaily:2016mnp}
J.~L. Bourjaily, S.~Franco, D.~Galloni, and C.~Wen, ``{Stratifying On-Shell
  Cluster Varieties: the Geometry of Non-Planar On-Shell Diagrams},''
  \href{http://dx.doi.org/10.1007/JHEP10(2016)003}{{\em JHEP} {\bfseries 10}
  (2016) 003},
\href{http://arxiv.org/abs/1607.01781}{{\ttfamily arXiv:1607.01781 [hep-th]}}.
%%CITATION = ARXIV:1607.01781;%%.

\bibitem{Bern:2015ple}
Z.~Bern, E.~Herrmann, S.~Litsey, J.~Stankowicz, and J.~Trnka, ``{Evidence for a
  Nonplanar Amplituhedron},''
  \href{http://dx.doi.org/10.1007/JHEP06(2016)098}{{\em JHEP} {\bfseries 06}
  (2016) 098},
\href{http://arxiv.org/abs/1512.08591}{{\ttfamily arXiv:1512.08591 [hep-th]}}.
%%CITATION = ARXIV:1512.08591;%%.

\bibitem{Feng:2009ei}
B.~Feng, J.~Wang, Y.~Wang, and Z.~Zhang, ``{BCFW Recursion Relation with
  Nonzero Boundary Contribution},''
  \href{http://dx.doi.org/10.1007/JHEP01(2010)019}{{\em JHEP} {\bfseries 01}
  (2010) 019},
\href{http://arxiv.org/abs/0911.0301}{{\ttfamily arXiv:0911.0301 [hep-th]}}.
%%CITATION = ARXIV:0911.0301;%%.

\bibitem{Arkani-Hamed:2017jhn}
N.~Arkani-Hamed, T.-C. Huang, and Y.-t. Huang, ``{Scattering Amplitudes For All
  Masses and Spins},''
\href{http://arxiv.org/abs/1709.04891}{{\ttfamily arXiv:1709.04891 [hep-th]}}.
%%CITATION = ARXIV:1709.04891;%%.

\bibitem{Huang:2013owa}
Y.-T. Huang and C.~Wen, ``{ABJM amplitudes and the positive orthogonal
  grassmannian},'' \href{http://dx.doi.org/10.1007/JHEP02(2014)104}{{\em JHEP}
  {\bfseries 02} (2014) 104},
\href{http://arxiv.org/abs/1309.3252}{{\ttfamily arXiv:1309.3252 [hep-th]}}.
%%CITATION = ARXIV:1309.3252;%%.

\bibitem{Huang:2014xza}
Y.-t. Huang, C.~Wen, and D.~Xie, ``{The Positive orthogonal Grassmannian and
  loop amplitudes of ABJM},''
  \href{http://dx.doi.org/10.1088/1751-8113/47/47/474008}{{\em J. Phys.}
  {\bfseries A47} no.~47, (2014) 474008},
\href{http://arxiv.org/abs/1402.1479}{{\ttfamily arXiv:1402.1479 [hep-th]}}.
%%CITATION = ARXIV:1402.1479;%%.

\bibitem{Cachazo:2018hqa}
F.~Cachazo, A.~Guevara, M.~Heydeman, S.~Mizera, J.~H. Schwarz, and C.~Wen,
  ``{The S Matrix of 6D Super Yang-Mills and Maximal Supergravity from Rational
  Maps},''
\href{http://arxiv.org/abs/1805.11111}{{\ttfamily arXiv:1805.11111 [hep-th]}}.
%%CITATION = ARXIV:1805.11111;%%.

\bibitem{Bourjaily:2012gy}
J.~L. Bourjaily, ``{Positroids, Plabic Graphs, and Scattering Amplitudes in
  Mathematica},''
\href{http://arxiv.org/abs/1212.6974}{{\ttfamily arXiv:1212.6974 [hep-th]}}.
%%CITATION = ARXIV:1212.6974;%%.

\end{thebibliography}\endgroup

\end{document}